%% file: main.tex
\documentclass[runningheads]{llncs}
\usepackage{url} 
\usepackage[ruled,lined]{algorithm2e}
\usepackage{graphicx}

\usepackage{amssymb}
\usepackage{amsmath}
\usepackage{array}
\usepackage{enumitem}

\usepackage{nameref}
\usepackage{hyperref} 
\usepackage{subcaption} 
\usepackage{mathtools} 
\usepackage{centernot}
\usepackage{float}

\usepackage{graphicx,booktabs}
\usepackage{multirow}
\usepackage[figuresleft]{rotating}

\def\mi#1{\mathit{#1}}
\def\univ{{\cal U}}

\def\Uact{\univ_{\mi{act}}}

\def\bag{\mathcal{B}}
\def\la{\langle}
\def\ra{\rangle}

\def\UAN{\univ_{\mi{AN}}}

\def\post#1{\ensuremath{{#1}\kern-.05ex\bullet}}

\makeatletter
\newcommand*{\xxRightarrow}[2][]{%
  \ext@arrow 0359\xxRightarrowfill@{#1}{#2}%
}
\newcommand*{\xxnRightarrow}[2][]{%
  \ext@arrow 0359\xxRightarrowfill@{#1}{{\raisebox{-0.4ex}[0ex][0ex]{\tiny$#2$}}}%
}
\newcommand*{\xxRightarrowfill@}{%
  \narrowfill@\Relbar\Relbar\Rightarrow{\vcenter{\hbox{\scalebox{0.7}{$\diagup$}}}}
}
\newcommand*{\narrowfill@}[5]{%
  $\m@th\thickmuskip0mu\medmuskip\thickmuskip\thinmuskip\thickmuskip
  \relax#5#1\mkern-7mu%
  \cleaders\hbox{$#5\mkern-2mu#2\mkern-2mu$}\hfill
  \mkern-5mu %
  #4%
  \mkern-5mu %
  \cleaders\hbox{$#5\mkern-2mu#2\mkern-2mu$}\hfill
  \mkern-7mu#3$%
}
\makeatother
\newcommand\actend{\vcenter{\hbox{\scalebox{0.75}{$\blacksquare$}}}}
\newcommand\actstart{\vcenter{\hbox{\scalebox{0.75}{$\blacktriangleright$}}}}

\newcommand\actStartAndEnd{\{ \actstart, \actend\}}

\newcommand\loopActs{\mathcal{A}_{loop}}
\newcommand\skipActs{\mathcal{A}_{skip}}
\newcommand\loopAct[2]{\ensuremath{\mathrm{loop}_{#1,#2}}}
\newcommand\skipAct[2]{\ensuremath{\mathrm{skip_{#1,#2}}}}
\newcommand\logActs{act(L)}
\newcommand\repairedLog{\hat{L}}
\newcommand\repairedLogActs{act(\repairedLog)}
\newcommand\remainingLogActs{\mathcal{A}_{L}}
\newcommand\multiO{\left[\,}
\newcommand\multiC{\,\right]}
\newcommand\cxRightarrow[2]{%
  \xRightarrow[#1]{\raisebox{-0.3ex}[0ex][0ex]{\tiny$#2$}}
}
\newcommand\dfVal[3]{{\cxRightarrow{}{#3}}(#1,#2)}
\newcommand\df[4]{#1 \cxRightarrow{#4}{#3} #2}
\newcommand\Sdf[4]{\smash{#1 \cxRightarrow{#4}{#3} #2}}
\newcommand\SdfVal[3]{\smash{{\cxRightarrow{}{#3}}(#1,#2)}}
\newcommand\nDf[4]{#1 \xxnRightarrow[#4]{#3} #2}

\DeclareMathOperator{\cupdot}{\dot{\cup}\,}

\begin{document}
\title{Revisiting the Alpha Algorithm To Enable Real-Life Process Discovery Applications -- Extended Report}
\titlerunning{Revisiting the Alpha Algorithm}

\author{Aaron Küsters\orcidID{0009-0006-9195-5380}, Wil M.P. van der Aalst\orcidID{0000-0002-0955-6940}}
\authorrunning{Aaron Küsters, Wil van der Aalst}
%
\institute{Process and Data Science (PADS), RWTH Aachen University, Germany\\
\href{mailto:kuesters@pads.rwth-aachen.de}{\texttt{kuesters@pads.rwth-aachen.de}} 
\\
\href{mailto:wvdaalst@pads.rwth-aachen.de}{\texttt{wvdaalst@pads.rwth-aachen.de}}
 }
\maketitle              

\begin{abstract}
The Alpha algorithm was the first process discovery algorithm that was able to discover process models with concurrency based on incomplete event data while still providing formal guarantees. However, as was stated in the original paper, practical applicability is limited when dealing with exceptional behavior and processes that cannot be described as a structured workflow net without short loops.
This paper presents the Alpha+++ algorithm that overcomes many of these limitations, making the algorithm competitive with more recent process mining approaches.
The different steps provide insights into the practical challenges of learning process models with concurrency, choices, sequences, loops, and skipping from event data. The approach was implemented in ProM and tested on various publicly available, real-life event logs.

\end{abstract}

\section{Introduction}
\label{sec:intro}
\input{intro.tex}

\section{Preliminaries}\label{sec:prelims}
\input{prelims.tex}
\section{Alpha+++}\label{sec:newalpha}
\input{approach.tex}

\section{Implementation}\label{sec:impl}
\input{implementation.tex}
\section{Evaluation}\label{sec:eval}
\input{evaluation.tex}
\section{Conclusion}\label{sec:concl}
\input{concl}

\clearpage
\bibliographystyle{plain}
\bibliography{lit,other}
\appendix
    \label{appendix}
    \input{appendix.tex}

\end{document}

%% file: intro.tex
The original \emph{Alpha algorithm} was developed over twenty years ago \cite{aal_edcis,aal_min_TKDE}. The goal of the algorithm was to show the challenges related to discovering process models with concurrency from example traces. It was formally proven that, a process modeled as a structured workflow net without short loops, can be rediscovered from an event log that is directly-follows complete \cite{aal_min_TKDE}. 
Despite this remarkable theoretical result, the Alpha algorithm has limited practical relevance for two main reasons:
\begin{itemize}
    \item The original algorithm did not attempt to filter out infrequent behavior. Since exceptional behavior is not separated from frequent behavior, it is generally impossible to uncover structure from real-life event logs.
    \item The original algorithm assumed that the process can be modeled as a free-choice Petri net with unique visible activity labels. Most real-life processes can \emph{not} be modeled as a structured workflow net without short loops and unique visible labels.
\end{itemize}
These limitations were already acknowledged in the papers proposing the algorithm, e.g., 
the focus of \cite{aal_min_TKDE} was on showing the theoretical limits of process discovery based on directly-follows complete event logs.
Many of the later process discovery approaches use these insights.
Various extensions of the Alpha algorithm have been proposed, e.g., 
\cite{wen_nfc-min_DMKD} extends the core algorithm to deal with long-term dependencies, and
\cite{wen_inv-task-dke} extends the core algorithm to deal with invisible activities (e.g., skipping).
Region-based process-discovery approaches provide 
formal guarantees.
State-based regions were introduced by Ehrenfeucht and Rozenberg \cite{ehrenfeucht_regions} in 1989
and generalized by Cortadella et al.\ \cite{Cortadella98}.
In \cite{two-step-mining-SOSYM}, it is shown how these state-based regions 
can be applied to process mining by first creating a log-based automaton using different abstractions.
In \cite{DBLP:conf/bpm/CarmonaCK08,carmona-PN2010}, refinements are proposed to tailor state-based regions toward process discovery.
Language-based regions work directly on traces without creating an automaton first; see, for example, the approaches presented in
\cite{lorenz_BPM2007,language_mining_dongen-fundamenta2009,bas-ilp-computing}.

Variants of the Alpha algorithm and the region-based approaches have problems dealing with infrequent behavior and are rarely used in practice. The region-based approaches are also infeasible for larger models and logs. Approaches such as the eST-Miner \cite{lisa-lt-dep-PN2022} and the different variants of the inductive miner \cite{sander-infreq-bpi2013-lnbip2014,sander-scalable-procmin-SOSYM} aim to provide formal guarantees but can also handle infrequent behavior.
Variants of the inductive miner have also been implemented in various commercial systems (e.g., Celonis).
The so-called split-miner uses a combination of approaches to balance recall and precision \cite{split-miner}.

The goal of this paper is to go back to the original ideas used by the Alpha algorithm and make the algorithm work in practical settings. The result is the \emph{Alpha+++ algorithm}, which, not only extends the core algorithm, but also removes problematic noisy activities, adds invisible activities, repairs loops, and post-processes the resulting Petri net.
The approach uses a broad combination of novel ideas, making the Alpha algorithm competitive when compared with the state-of-the-art. The ideas incorporated in the Alpha+++ algorithm may also be used in combination with other approaches (e.g., identifying problematic activities and introducing artificially created invisible activities).

The remainder of this paper is organized as follows.
Section~\ref{sec:prelims} introduces event logs, directly-follows graphs, and the original Alpha algorithm. Section~\ref{sec:newalpha} describes the Alpha+++ algorithm.
The algorithm has been implemented in ProM (cf.\ Section~\ref{sec:impl}) 
and evaluated using various event logs (cf.\ Section~\ref{sec:eval}).
Section~\ref{sec:concl} concludes the paper.

%% file: prelims.tex
\subsection{Event Logs}

Process mining starts from event data. An event may have many different attributes.
However, here we focus on discovering the control flow and assume that each \emph{event} has a \emph{case} attribute, an \emph{activity} attribute, and a \emph{timestamp} attribute. 
We only use the timestamps to order events related to the same case.
Therefore, each case can be described as a sequence of activities, also called \emph{trace}. 
An \emph{event log} is a multiset of traces, as different cases can exhibit the same trace.

\begin{definition}[Event Log]\label{def:simplogrepeat}
  ~$\Uact$ is the universe of activity names. 
  A trace $\sigma = \la a_1,a_2,\allowbreak \ldots,\allowbreak a_n \ra \in {\Uact}^*$ is a sequence of activities.
  An event log $L \in \bag({\Uact}^*)$ is a multiset of traces.
  \end{definition}

For example, $L_1=[\la a,b,c,d \ra^{400},\langle a,b,d \rangle^{250}, \langle d,a,b,c \rangle^{4}, \langle d,a,b \rangle^{2} ]$ is an event log containing $656$ cases with $4$ different variants. Variant $\langle a,b,c,d\rangle$ is the most frequent one, i.e., $L_1(\langle a,b,c,d \rangle) = 400$.

We write $\mi{actMult}(L) = \multiO \sigma(i) \mid \sigma \in L \land 1 \leq i \leq |\sigma| \multiC$ for the multiset of activities in an event log $L$ and $act(L)=\{a \mid a \in \mi{actMult}(L)\}$ for the set of activities.

\subsection{Directly-Follows Graphs}
A \emph{Directly-Follows Graph} (DFG) is a graph showing how often one activity is followed by another.
A DFG consists of the activities as nodes and has an arc from an activity $a\in \Uact$ to an activity $b\in \Uact$ if $a$ is directly followed by $b$.
Two special nodes, corresponding to a start and an end node, are added additionally.

\begin{definition}[Directly-Follows Graph]\label{def:dfg}
  A Directly-Follows Graph (DFG) is a pair $G = (A,\smash{\cxRightarrow{}{G}})$,
  where $A \subseteq \Uact$ is a set of activities and
  $\smash{\cxRightarrow{}{G}} \in \bag((A \times A) \cup (\{\actstart\} \times A) \cup (A \times \{\actend\}) \cup (\{\actstart\} \times \{\actend\}))$ is a multiset of arcs.
  $\actstart$ is the start node and $\actend$ is the end node.
\end{definition}

Note that a DFG has arc weights. Hence, $\smash{\cxRightarrow{}{G}}$ is a multiset, where $\smash{\cxRightarrow{}{G}}(a,b)$ denotes how often $a$ is followed by $b$.
We write $\Sdf{a}{b}{G}{}$ if and only if $\SdfVal{a}{b}{G}{} > 0$ holds.
Similarly, we say that $\smash{\df{a}{b}{G}{\geq t}} \vspace{0.175cm}$ holds if and only if $\SdfVal{a}{b}{G}{} \geq t$. 

The construction of a DFG from an event log is straightforward.

\begin{definition}[Constructing DFGs from Event Logs]
 Let $L \in  \bag({\Uact}^*)$ be an event log.
 We can construct a DFG $\mi{disc}_{\mi{dfg}}(L) = (A, \smash{\cxRightarrow{}{L}})$ based on the directly-follows relations of event log $L$, with the set of activities $A = \{a \in \sigma \mid \sigma \in L\}$ and the multiset of arcs $\smash{\cxRightarrow{}{L}} = \left[(\sigma_i,\sigma_{i+1}) \mid \sigma \in L' \land \allowbreak 1  \leq i < |\sigma| \right]$
, where 
 $L'= \left [\langle \actstart \rangle \cdot \sigma \cdot \langle \actend \rangle \mid \sigma \in L \right ]$
 denotes the event log where artificial start and end activities have been added.
\end{definition}

Given an event log $L$, we can construct a DFG $\mi{disc}_{\mi{dfg}}(L)=(A,\smash{\cxRightarrow{}{L}})$, and in the context of $L$ refer to the directly-follows relations in $L$ represented by $\smash{\cxRightarrow{}{L}}$ directly.

  \begin{figure}[thb!]
    \begin{subfigure}[t]{0.45\textwidth}
      \centering
      \includegraphics[width=0.4\textwidth]{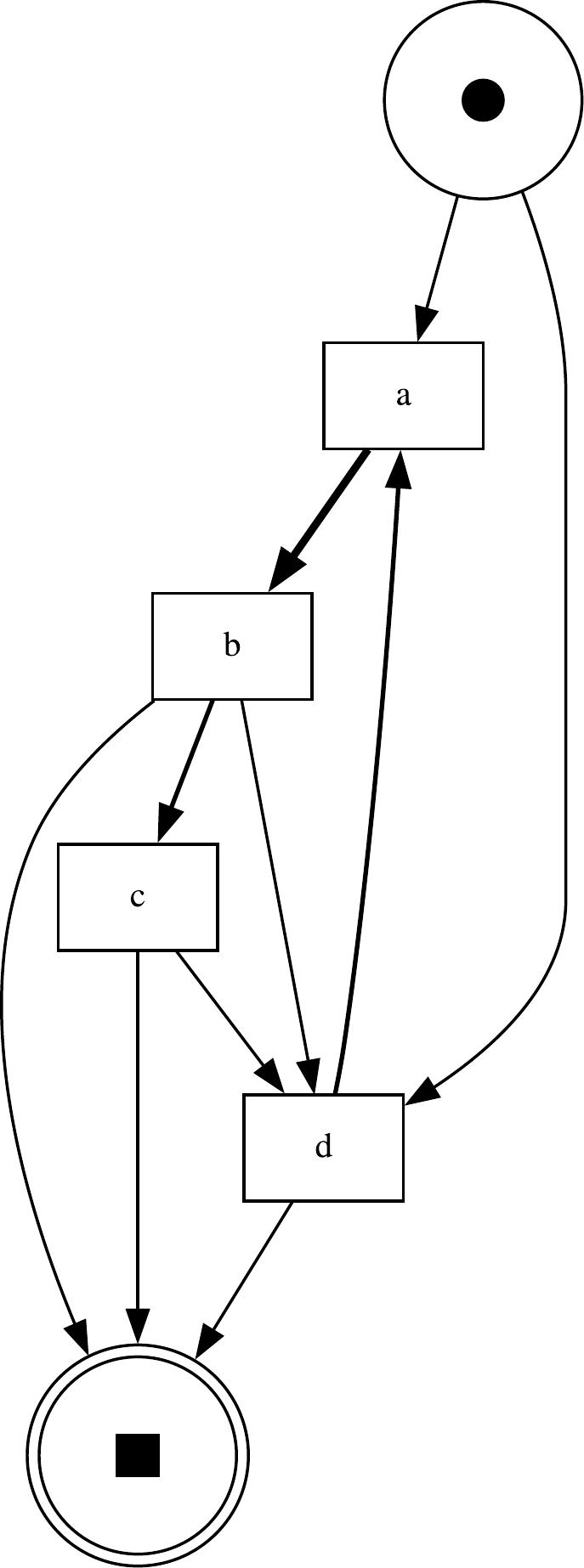}
      \caption{Basic DFG constructed from event log $L_1$.}
  \end{subfigure}%
   \hspace{1em}
  \begin{subfigure}[t]{0.45\textwidth}
    \centering
    \includegraphics[width=0.425\textwidth]{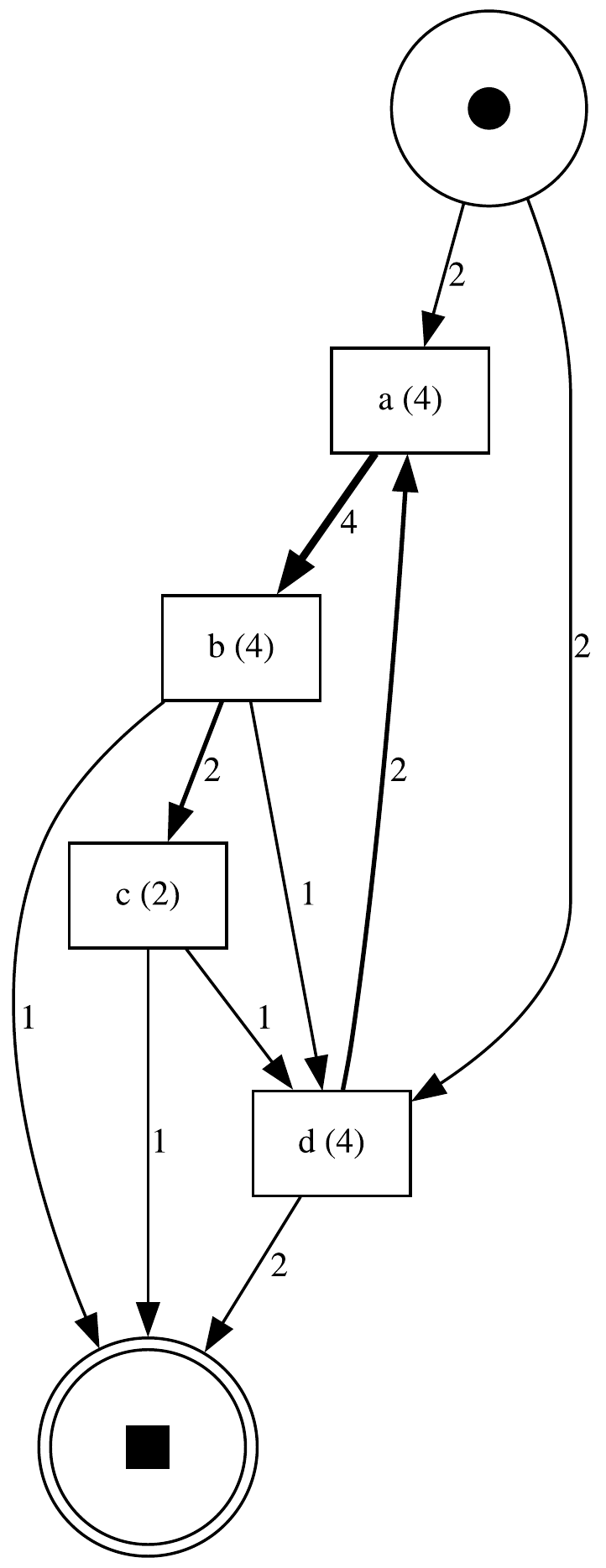}
    \caption{Weighted DFG for event log $L_1$ with annotated frequencies of the directly-follows relations and activities.}
\end{subfigure}%
    \centering
    \caption{Example DFGs for event log $L_1$.}
    \label{fig-prelim-dfg-example}
    \end{figure}

\subsection{Petri Nets}
We would like to discover process models which can represent more complex control-flow structures, like choices, loops, and concurrency. Therefore, we use \emph{labeled} Petri nets as a target format for process discovery. The reader is assumed to be familiar with the Petri net basics. 

  \begin{definition}[Labeled Petri Net]\label{def:lpn}
    A labeled Petri net is a tuple $N=(P,T,F,l)$ with a set of places $P$, a set of transitions $T$ (where $T \cap P = \emptyset$),
    a flow relation, $F\subseteq (P \times T) \cup (T \times P)$, and a labeling function $l \in T \not\rightarrow \Uact$.
    We write $l(t) = \tau$ if $t \in T \setminus \mi{dom}(l)$ (i.e., $t$ is a silent transition that cannot be observed).
    \end{definition}
    
 A marking is represented by a multiset of places $M \in \bag(P)$. For a node $x \in P \cup T$, we define the preset of $x$ as $\bullet x = \{y \in P \cup T \mid (y,x) \in F\}$ and the postset of $x$ as $x \bullet = \{y \in P \cup T \mid (x,y) \in F\}$.
 We focus on so-called \emph{accepting} Petri nets, i.e., Petri nets with a defined initial and final state.

    

  \begin{definition}[Accepting Petri Net]\label{def:apn}
    An accepting Petri net is a triplet $\mi{AN}=(N,\allowbreak M_{\mi{init}},\allowbreak M_{\mi{final}})$ where
    $N=(P,T,F,l)$ is a labeled Petri net,
    $M_{\mi{init}} \in \bag(P)$ is the initial marking, and $M_{\mi{final}} \in \bag(P)$ is the final marking.
    $\UAN$ is the set of accepting Petri nets.
    \end{definition}

The language defined by an accepting Petri net is then simply given by the set of traces corresponding to all firing sequences that start in the initial marking $M_{init}$ and end in the final marking $M_{final}$. A firing sequence leading from $M_{init}$ to $M_{final}$ is converted into a trace, i.e., a sequence of activities.
Note that transitions that fire are mapped onto the corresponding activities.
If a transition $t$ is silent (i.e., $l(t) = \tau$), no corresponding activity is created when firing $t$.
Hence, the language of an accepting Petri net is a set of traces. 

\subsection{Alpha Algorithm}
A \emph{process discovery algorithm} aims to discover a model from event data such that the language of the model best characterizes the example behavior seen in the event log. 
\begin{definition}[Process Discovery Algorithm]\label{def:pdalg}
  A process discovery algorithm is a function 
  $\mi{disc} \in \bag({\Uact}^*) \rightarrow \UAN$, i.e., 
  based on a multiset of traces, an accepting Petri net is discovered.
  \end{definition}

  The classical Alpha process discovery algorithm was introduced in~\cite{aal_min_TKDE}.
  To be able to better explain the extensions presented in this paper, we split
  the description into three main parts.
  From an input event log $L$, place candidates are constructed based on the directly-follows relations of the log.
  The resulting set of place candidates is pruned to remove dominated candidates.
  Finally, the discovered Petri net is constructed.
\begin{description}
\item[Candidate Building]
\begin{align*}
  \mi{Cnd} = \{ (A,B) \mid \emptyset \subsetneq A,B \subseteq act(L) &\land \forall_{a\in A}\forall_{b \in B} (\df{a}{b}{L}{})
  \\ &\land \forall_{a,a' \in A} (\nDf{a}{a'}{L}{})
   \land \forall_{b,b' \in B} (\nDf{b}{b'}{L}{}) \}
\end{align*}
\item[Candidate Pruning]
\begin{align*}
  \mi{Sel} = \{ (A_1,A_2)\in \mi{Cnd} \mid \forall_{(A_1',A_2')\in \mi{Cnd}}\; (( & A_1 \subseteq A_1' \land A_2 \subseteq A_2') \\ &\Rightarrow (A_1,A_2) = (A_1',A_2')) \}
\end{align*}

\item[Petri Net Construction] Let $PN=((P,T,F,l),M_{init},M_{final})$, where:
\begin{itemize}
  \item $P = \{p_{(A,B)} \mid (A,B) \in \mi{Sel}\} \cup \{i_W,o_W \}$
  \item $T = \{t_a \mid a\in act(L)\}$
  \item $F = \{ (t_a,p_{(A,B)}) \mid (A,B) \in \mi{Sel} \land a \in A\} \cup \{ (p_{(A,B)},t_b) \mid (A,B) \in \mi{Sel} \allowbreak \land  b \in B\} \cup \{ (i_W,t_s) \mid \exists_{\sigma} \ \langle s \rangle \cdot \sigma \in L\} \cup \{ (t_e,o_W) \mid \exists_{\sigma} \ \sigma \cdot  \langle e \rangle\in L\}$
  \item $l = \{(t_a,a) \mid a \in act(L)\} $
  \item $M_{init}=[i_W]$
  \item $M_{final}=[o_W]$
\end{itemize}
\end{description}

%% file: approach.tex
In this section, we introduce the Alpha+++ process discovery algorithm based on the classical Alpha algorithm.
Through certain pre-processing steps on the event log and a corresponding DFG, as well as fitness-based place filtering, this algorithm is especially well suited for real-life event logs.

The input for this process discovery algorithm is an event log $L$.
In particular, only ordered traces of activities with corresponding frequencies are required.
For the main steps of the algorithm, a DFG based on the event log $L$ is used exclusively.
Traces of the event log are only used for replay to remove unfitting place candidates.
For simplicity, we assume that the traces of $L$ already include artificial start and end activities, in particular, we assume $\actStartAndEnd \subseteq \logActs{}$.

We introduce the steps of the algorithm in the following order:
\begin{enumerate}
  \item \emph{Determine Activities}, where the set of activities used throughout the algorithm is determined. Problematic activities are removed from the event log and artificial activities are added, resulting in a repaired event log $\repairedLog$.
  \item \emph{Create an Advising DFG}, where an \emph{advising DFG} is constructed based on the DFG corresponding to the repaired log $\repairedLog$, retaining only some of the original DFG edges.
  \item \emph{Candidate Building}, where a set of place candidates is built based on the directly-follows relation of the activities.
  \item \emph{Candidate Pruning}, where through efficient multistep filtering unfit or undesirable place candidates are discarded.
  \item \emph{Petri Net Construction}, where a Petri net is constructed based on the activities of the event log, the added artificial activities and the remaining place candidates.
  \item \emph{Post-Processing Petri Net}, where the repaired event log is replayed on the Petri net to remove problematic places.
\end{enumerate}
\subsection{Determine Activities}
First, we determine the set of activities used in the later steps.
Initially, starting with the set of activities occurring in the log, we first remove problematic activities that can cause issues with discovering place candidates later on.
Next, we also add artificial activities to allow discovery of place candidates for certain loop and skip constructs.

\subsubsection{Removing Problematic Activities}
Problematic activities can significantly alter the directly-follows relations of an event log, which are used in the later steps to identify place candidates.
In the most extreme case, if a problematic activity randomly occurs between any other two activities in all traces, all the directly-follows information between two other activities would be lost.

We select a subset $\remainingLogActs \subseteq act(L)$ of activities to keep and remove the other problematic activities $act(L) \setminus \remainingLogActs$.
There are many possible approaches to identifying problematic activities, such as calculating a problem-score per activity and considering all values above a certain threshold as problematic.
For instance, for a simple problem-score, the fraction of directly-follows relation involving an activity which are parallel, i.e., also occur in the opposite direction, could be considered.
This would, for example, allow to correctly identify the problem in the aforementioned extreme case.

\subsubsection{Adding Artificial Activities}
Discovering Petri net constructs involving silent transitions is a non-trivial task for a DFG-based algorithm.
Additionally, in later steps, we want to use traces from the log to assess the fitness of place candidates.
Silent activities make calculating fitness scores significantly harder, as then token-based replay is no longer sufficient and computationally expensive alignments have to be computed.
As a solution, we propose adding \emph{artificial activities} to traces.
They are not part of the activity set of the event log and are only used to find and evaluate place candidates.
In the final discovered Petri net, these artificial activities are then translated as silent transitions.
This allows discovering Petri nets with silent transitions, while still retaining the advantages of token-based replay fitness evaluation during the algorithm steps.
We add artificial activities for two types of constructs: \emph{Loops} and \emph{Skips}.

\begin{figure}[tbp]
  \centering
  \includegraphics[width=1.0\textwidth]{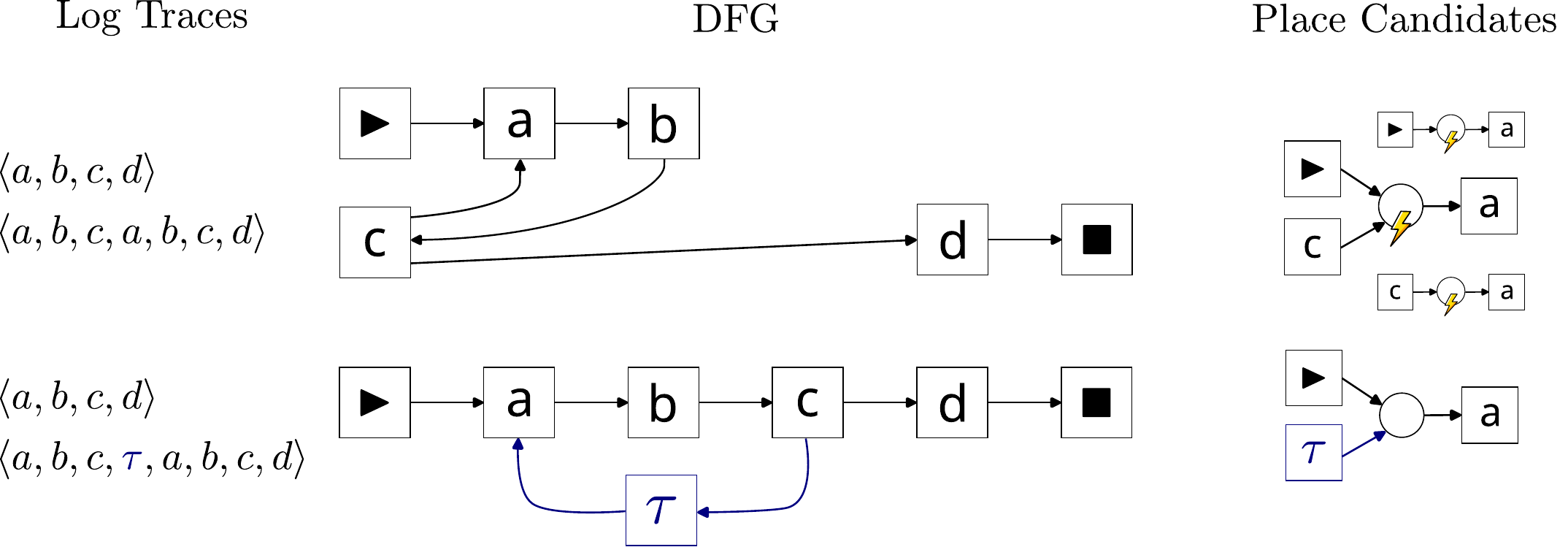}
  \caption{Two event logs (traces shown on the left) and their DFGs.
           In the first event log ($L_\circlearrowleft$) the directly-follows relation between $\actstart$ and $a$ is the same as between $c$ and $a$.
           This causes issues, as the corresponding place candidates all have very low fitness.
           The added artificial activity $\tau$ inserted between the looped sequence $\langle a,b,c \rangle$ solves this problem, as the problematic directly-follows relation between $c$ and $a$ is replaced.
         }
  \label{fig-main-loop-repair}
\end{figure}

Adding artificial activities for loops is necessary, as the directly-follows relation between an end activity and a start activity of a loop can cause the discovery of problematic places.
For example, consider the event log $L_\circlearrowleft=[\langle a,b,c,d \rangle, \allowbreak \langle a,b,c,a,b,c,d \rangle]$.
Clearly, this event log can be nicely expressed by a Petri net containing a loop construct, which allows repeating the activities $a,b,c$.
However, the directly-follows relation $\smash{\df{c}{a}{L_\circlearrowleft}{}}$ prevents discovering this loop accurately, as shown in \autoref{fig-main-loop-repair}.

We detect loop constructs based on the directly-follows relations of the input event log $L$.
For a given threshold $d \in \mathbb{R}^+$, we can define the set of detected loops:
\begin{definition}[Detected Loops]
  Let $\mi{loops}$ be the function that maps an event log to the set of detected loop start and end activities.
  \begin{align*}
    \mi{loops}(L)=\{(b,a) \in \logActs \times \logActs
    \mid~&\exists_{(x_1,\dots,x_k) \in \logActs^*, i \in \{1,\dots,k\}}(x_i = a
    \\&\land \forall_{i\in\{1,\dots,k-1\}}(\df{x_i}{x_{i+1}}{L}{\geq d}) \\
    &\land \df{x_k}{b}{L}{\geq d} \land \df{b}{a}{L}{\geq d})\}  
   \end{align*}
\end{definition}

The parameter $d$ determines the minimal DFG edge weight to consider when looking for loops.
For example, with a threshold $d{=}1$ and the event log $L_\circlearrowleft$, we can calculate $\mi{loops}(L_\circlearrowleft) = \{(c,a)\}$.
As loop constructs can make a process model very imprecise, we do not want to falsely detect loop behavior from rather infrequent directly-follows relations.
For convenience, we can also consider threshold values $d$ relative to the mean directly-follows weight. 

For each detected loop endpoint pair $(b,a)\in \mi{loops}(L)$, we want to add an artificial activity $\loopAct{b}{a}\not\in \logActs$.
We write $\loopActs = \{ \loopAct{b}{a} \mid (b,a) \in \mi{loops}(L)  \}$ to denote the set of added artificial loop activities.
Additionally, we define a transformation function which transforms a trace $\sigma \in L$ to a trace $\sigma' \in (\remainingLogActs \cup \loopActs)^*$.

\begin{definition}[Loop Repair Function]
  Let $\mi{repair}_\circlearrowleft$ be the function that transforms a trace $\sigma$ into a repaired trace with added artificial loop activities.
  \begin{align*}
    \mi{repair}_\circlearrowleft(\sigma,L) =  \begin{cases}
      \langle b, \loopAct{b}{a}, a \rangle \cdot \mi{repair}_\circlearrowleft(\sigma',L) & \mathrm{ if }\; \exists_{(b,a)\in \mi{loops}(L)}\sigma = \langle b, a \rangle \cdot \sigma'\\
      \langle \rangle &  \mathrm{ if }\; \sigma=\langle \rangle \\
            \langle x \rangle \cdot \mi{repair}_\circlearrowleft(\sigma',L) &  \mathrm{ otherwise, with }\; \sigma=\langle x \rangle \cdot \sigma' \\
    \end{cases}
  \end{align*}
\end{definition}
This function will be later used to transform the input event log $L$ into a repaired event log, in which artificial activities have been added to relevant traces.

\begin{figure}[hbt]
  \centering
  \includegraphics[width=1.0\textwidth]{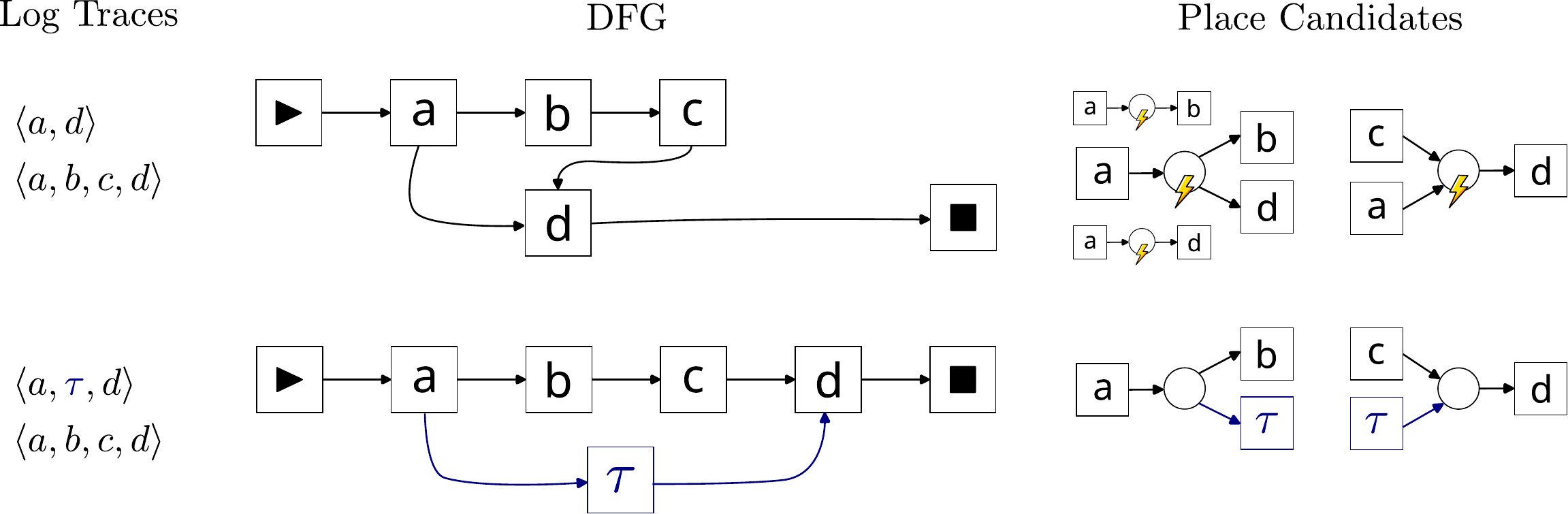}
  \caption{
Two event logs and their DFGs showcasing the motivation for repairing implicit skips.
         The directly-follows relation between $a$ and $d$ would suggest considering place candidates with poor fitness.
           The second log, where an artificial activity $\tau$ is inserted where $b$ and $c$ are skipped, mitigates this problem by replacing the directly-follows relation between $a$ and $d$.
         }
  \label{fig-main-skip-repair}
\end{figure}

Next, we describe how artificial activities can assist in correctly discovering activity \emph{Skips}, as shown in \autoref{fig-main-skip-repair}.

For a directly-follows-weight threshold $d \in \mathbb{R}^+$, the detected skips for event log $L$ are defined by the following function, which provides the set of activities that have been detected as being ``skippable'' after an activity $a\in (\remainingLogActs \cup \loopActs)$.

\begin{align*}
\mi{skips}(a,L) = \{ b \in \logActs{} &\mid~\df{a}{b}{L}{} \land \nDf{a}{a}{L}{} \land \nDf{b}{a}{L}{\geq d} \land  \nDf{b}{b}{L}{\geq d} \land a,b \not\in \{\actstart,\actend\} \\
 &\land \emptyset \subsetneq \{x\in \logActs{} \mid \df{b}{x}{L}{\geq d}\} \subseteq \{x\in \logActs{} \mid \df{a}{x\}}{L}{\geq d}\} 
\end{align*}
If $B \in \mi{skips}(a,L)$ we assume that all activities $b \in B$ are optional steps after $a$.
To allow appropriate model discovery in the rest of the algorithm, the log is repaired using a new artificial activity $skip_{a,B} \not\in \logActs$.
The set of all artificial skip activities is denoted by $\skipActs$.
This artificial skip activity is inserted everywhere in a trace $\sigma$ of $L$, where activity $a$ is not directly followed by an activity $b \in B$ in $\sigma$ (i.e.,~$b$ was skipped). 
For that, we define the following transformation function:
\begin{align*}
  \mi{repair}_\tau(\sigma,L) =  \begin{cases}
    \langle \rangle & \mathrm{ if }\; \sigma = \langle \rangle \\
    \langle x \rangle \cdot  \mi{repair}_\tau(\sigma',L) & \mathrm{ if }\; \sigma = \langle x \rangle \cdot \sigma' \land \mi{skips}(x,L) = \emptyset \\
    \langle a, \skipAct{a}{B} \rangle \cdot  \mi{repair}_\tau(\langle x \rangle \cdot \sigma',L) & \mathrm{ if }\; \sigma = \langle a, x \rangle \cdot \sigma' \land x\not \in \mi{skips}(a,L) \\
    \langle a, x \rangle \cdot  \mi{repair}_\tau(\sigma',L) & \mathrm{ if }\; \sigma = \langle a, x \rangle \cdot \sigma' \land x \in \mi{skips}(a,L) \\
  \end{cases}
  \end{align*}

We can now construct a \emph{repaired} event log $\repairedLog$ from the input event log $L$ based on the previously identified set of detected loops $\mi{loops}(L)$ and skips $\mi{skips}(L)$.
For that, we use their corresponding artificial activity set $\loopActs$ and $\skipActs$ as well as their corresponding trace transformation functions $ \mi{repair}_\circlearrowleft$ and $ \mi{repair}_\tau$ to transform the input event log $L$ into a repaired event log $\repairedLog$.
Note that $\repairedLogActs=\left(\remainingLogActs\cupdot \loopActs \cupdot \skipActs \right)$.

\[
  \repairedLog = [\mi{repair}_\tau\left(\mi{repair}_\circlearrowleft(\sigma,L {\upharpoonright} \remainingLogActs),L {\upharpoonright} \remainingLogActs\right)
   \mid \sigma\in L {\upharpoonright} \remainingLogActs]
\]

\subsection{Create an Advising DFG}
Next, we extract a pruned DFG from the repaired event log $\repairedLog$, which ignores infrequent directly-follows relations.
This DFG is used as guidance using the following algorithm steps.
Note that this step does not modify the repaired event log: The output of this step is a pruned DFG containing the activities $\repairedLogActs$ as nodes.
Edges between activities $a$ and $b$ are retained if their weight corresponds to at least $1\%$ of the sum of the weights of all incoming edges to $b$ or $1\%$ of the sum of all outgoing edges from $a$.
The value of $1\%$ was determined as a good cutoff through experimentation.
In addition, edges with weights below an absolute threshold value $n\in\mathbb{N}_0$ are also removed. 

For the repaired event log $\repairedLog$ and a given DFG-weight threshold $n\in\mathbb{N}_0$, we define the \emph{advising DFG} (abbreviated as aDFG) as follows:

\begin{align*}
   \mi{minW}(a,b) = 0.01\cdot\min\biggl\{\sum_{c\in \repairedLogActs}\dfVal{c}{b}{\repairedLog},\sum_{c\in \repairedLogActs}\dfVal{a}{c}{\repairedLog} \biggr\}
  \\
  \mathrm{aDFG} = \biggl(\repairedLogActs,\biggl [(a,b)\in \repairedLogActs^2 \bigg| \dfVal{a}{b}{\repairedLog} \geq \max\left\{n,\mi{minW}(a,b)\right\}
  \biggr ]\biggr)
\end{align*} 

\subsection{Candidate Building}
With the repaired event log and the aDFG, we can continue with building place candidates.
Place candidates are composed of two sets of activities: The first set corresponds to the transitions that should add a token to this place in a Petri net.
The second set corresponds to transitions that should remove a token from this place.

The set of all place candidates is given by:
\begin{align*}
  \mi{Cnd}_0 = \{ (A_1,A_2) \mid A_1,A_2 \subseteq \repairedLogActs & \land \forall_{a_1\in A_1}\forall_{a_2 \in A_2} (\df{a_1}{a_2}{\mathrm{aDFG}}{})\\
                                                                   & \land \forall_{a_1\in A_1} \forall_{a_2 \in A_1 \setminus A_2} (\nDf{a_1}{a_2}{\mathrm{aDFG}}{})   \\
                                                                   & \land \forall_{a_1\in A_2 \setminus A_1} \forall_{a_2 \in A_2} (\nDf{a_1}{a_2}{\mathrm{aDFG}}{})
                                                                   \\
                                                                    & \land \exists_{a_1 \in A_1 \setminus A_2} \exists_{a_2 \in A_2 \setminus A_1} (\nDf{a_2}{a_1}{\mathrm{aDFG}}{})
                                                                   \} \\
\end{align*}

\subsection{Candidate Pruning}
The set of place candidates $\mi{Cnd}_0$ includes many unfit places, which would produce process models with very low fitness.
Furthermore, some place candidates might be dominated by others (e.g., the place candidate $(\{a\},\{f\})$ is dominated by the candidate $(\{a,b\},\{e,f\})$).
Pruning the set of place candidates requires an efficient approach, as the number of place candidates can easily grow huge.
We propose a three-step pruning approach.
First, place candidates are filtered purely based on activity counts.
If the difference in frequency of the input and output activity set is relatively large, the place candidate is rather unfit.
This condition can be checked very efficiently.
Next, the local fitness of the place candidate is calculated based on local trace replay.
Local trace replay takes the order of the activities in the traces into account, and thus can detect even more unfit place candidates.
Finally, to remove dominated place candidates, we retain only maximal place candidates.
\subsubsection{Balance-based Pruning:}
For the balance-based pruning, we consider the number of activity occurrences in the log $\repairedLog$ using $\mi{actMult}(\repairedLog)$.
For a set of activities, $A\subseteq \repairedLogActs$ we can then sum the frequencies together as $\mi{count}(\repairedLog{},A)= \sum_{a\in A} \mi{actMult}(\repairedLog)(a)$.
Based on that, we define the $\mi{balance}$ of a candidate $(A_1,A_2)$:
\begin{align*}
  \mi{balance}(\repairedLog{},A_1,A_2) = \frac{|\mi{count}(\repairedLog{},A_1) - \mi{count}(\repairedLog{},A_2)|}{\max\{\mi{count}(\repairedLog{},A_1), \mi{count}(\repairedLog{},A_2)\}}
\end{align*}

The balance of a candidate is between $0$ and $1$.
Higher values are an indication that the place candidate is unfit.
Based on a balance threshold $b\in [0,1]$, candidates with a higher balance value than $b$ can be filtered out:
\begin{align*}
  \mi{Cnd}_1 = \{ (A_1,A_2)\in \mi{Cnd}_0 \mid \mi{balance}(\repairedLog{},A_1,A_2) \leq b \}
\end{align*}
\subsubsection{Fitness-based Pruning:}
\label{sssec:fitness-pruning}
Let $\mi{fit}(\sigma,(A_1,A_2),k)$ be defined as follows:
\begin{align*}
  \mi{fit}(\sigma,(A_1,A_2),k) =
  \begin{cases}
    1                          & \mathrm{if}\; \sigma = \langle \rangle, k=0                               \\
    0                          & \mathrm{if}\; \sigma = \langle \rangle, k\neq0                            \\
    0                          & \mathrm{if}\; \sigma = \langle a \rangle \cdot \sigma', k=0, a \not\in A_1, a \in A_2     \\
    \mi{fit}(\sigma',(A_1,A_2),k+1) & \mathrm{if}\; \sigma = \langle a \rangle \cdot \sigma', a \in A_1, a \not\in A_2          \\
    \mi{fit}(\sigma',(A_1,A_2),k-1) & \mathrm{if}\; \sigma = \langle a \rangle \cdot \sigma', k\geq 1, a \not\in A_1, a \in A_2 \\
    \mi{fit}(\sigma',(A_1,A_2),k)   & \mathrm{if}\; \sigma = \langle a \rangle \cdot \sigma', (a \in A_1 \cap A_2 \lor a \not\in A_1 \cup A_2)              \\
  \end{cases}
\end{align*}
Note that $\mi{fit}(\sigma,(A_1,A_2),0) = 1$ if the place candidate $(A_1,A_2)$ fits the trace; otherwise it takes the value $0$.

The traces relevant for a place candidate $(A_1,A_2)$ are defined by the following function:
\[
  \mi{rel}(A_1,A_2) = \left[ \sigma = \langle a_1,\dots,a_n \rangle \in \repairedLog \mid \exists_{i\in \{1,\dots,n\}} (a_i \in A_1 \lor a_i \in A_2)\right]
\]
We consider traces relevant for a place candidate, if they contain at least one activity that is in the set of outgoing or ingoing activities of that place candidate.
For a single activity, we use the notation $\mi{rel}(a):=\mi{rel}(\{a\}, \emptyset)$ to denote the traces containing that activity.

We write $\mi{fit}(\sigma,(A_1,A_2)) \coloneqq \mi{fit}(\sigma, (A_1,A_2),0)$ and
\[\mi{fit}(\repairedLog{},(A_1,A_2)) \coloneqq \sum_{\sigma \in \mi{rel}(A_1,A_2)} \mi{fit}(\sigma,(A_1,A_2))\]
for ease of notation.

For a given local candidate fitness threshold $t\in[0,1]$, the candidates remaining after the local fitness replay pruning are then given as:
\begin{align*}
  \mi{mfit}(A_1,A_2) = \min \left\{\frac{\sum_{\sigma \in \mi{rel}(a)} \mi{fit}(\sigma,(A_1,A_2))}{|\mi{rel}(a)|} \,\bigg|\, a\in A_1 \cup A_2 \right\} \\
  \mi{Cnd}_2 = \left\{ (A_1,A_2)\in \mi{Cnd}_1 \,\bigg|\, \frac{\mi{fit}(\repairedLog{},(A_1,A_2))}{|\mi{rel}(A_1,A_2)|} \geq t \land \mi{mfit}(A_1,A_2) \geq t \right\}
\end{align*}
 
\subsubsection{Maximal Candidate Selection:}
Finally, as the last candidate pruning step, all dominated place candidates are removed, just like in the original Alpha algorithm.
\begin{align*}
  \mi{Sel} = \{ (A_1,A_2)\in \mi{Cnd}_2 \mid \forall_{(A_1',A_2')\in \mi{Cnd}_2}\; (( & A_1 \subseteq A_1' \land A_2 \subseteq A_2') \\ &\Rightarrow (A_1,A_2) = (A_1',A_2')) \}
\end{align*}

\subsection{Petri Net Construction}
Based on the remaining place candidates, an accepting Petri net is constructed as the tuple $((P,T,F,l),M_{init},M_{final})$, where
\begin{itemize}
  \item $P = \{p_{(A_1,A_2)} \mid{} (A_1,A_2) \in \mi{Sel}\}$
  \item $T = \{t_a \mid{} a\in \repairedLogActs \setminus \actStartAndEnd \}$
  \item $F = \{ (t_a,p_{(A_1,A_2)}) \mid{} (A_1,A_2) \in \mi{Sel} \land a \in A_1 \setminus \actStartAndEnd \} \cup \{ (p_{(A_1,A_2)},t_a) \mid (A_1,A_2) \in \mi{Sel} \land a \in A_2 \setminus \actStartAndEnd\} $
  \item $l = \{(t_a,a) \mid{} a \in \remainingLogActs{}\}\cup \{(t_a,\tau) \mid a \in (\loopActs \cup \skipActs \}$
  \item $M_{init}=\multiO p_{(A_1,A_2)} \in P \mid{} \actstart \in A_1 \multiC$
  \item $M_{final}=\multiO p_{(A_1,A_2)} \in P \mid{} \actend \in A_2 \multiC$
\end{itemize}
are the components defined using the results of the previous steps.
\subsection{Post-Processing Petri Net} 
Let $\mi{replay}(p,PN,\sigma)$ be the replay function, which takes the value $1$ exactly when the place $p$ of the Petri net $PN$ can replay trace $\sigma$ (i.e., there is no missing or remaining token in $p$ at any time when replaying $\sigma$ on $PN$).

For a given local place replay fitness threshold $r\in [0,1]$, we can then define the result of the post-process replay as $((P',T,F',l),M_{init}',M_{final}')$, where the set of updated places $P'$ is given by:
 \[
  P' = \left\{p_{(A_1,A_2)} \mid{} (A_1,A_2) \in \mi{Sel} \land \frac{\sum_{\sigma \in rel(A_1,A_2)} \mi{replay}(p,PN,\sigma)}{|rel(A_1,A_2)|}\geq r\right\}
  \]
  The flow relation and initial and final markings are also updated correspondingly:
\begin{itemize}
  \item $F' = \{(i,o) \in F \mid i\in P' \land o\in P'\}$
  \item $M_{init}'= \multiO p \in  M_{init} \mid p \in P' \multiC$
  \item $M_{final}'= \multiO p \in M_{final} \mid p \in P' \multiC$
  \end{itemize}

  Note, that this post-processing uses a similar local fitness measure as the fitness-based pruning presented in \autoref{sssec:fitness-pruning}.
  One key difference is the handling of self-loops (where the post-processing replay is more restrictive).
  Another aspect to consider is that the candidate fitness pruning occurs before deleting dominated place candidates, while the post-processing fitness replay occurs after this pruning.
  In particular, choosing more aggressive filtering in the post-processing can lead to a decrease in places and an increase in disconnected transitions, which is often undesirable.

  The final accepting Petri net discovered is $((P',T,F',l),M_{init}',M_{final}')$.

%% file: implementation.tex
We implemented the Alpha+++ algorithm as a ProM\footnote{\url{https://promtools.org/}} plugin (Java) and also created a Python implementation\footnote{\url{https://github.com/aarkue/alpha-revisit-python}} for large-scale evaluation on a variety of real-life event logs.
The ProM plugin (AlphaRevisitExperiments\footnote{\url{https://svn.win.tue.nl/repos/prom/Packages/AlphaRevisitExperiments/}}
) can be installed in ProM Nightly versions and can be used in standard mode to simply discover a Petri net or in interactive mode to experiment with different algorithm step options and view additional information (e.g., how many place candidates were pruned in which step).
In both versions, the Alpha+++ preset can be selected out of the preset list on the top.
The parameters used throughout the algorithm steps can then be changed.
Additionally, the different algorithm steps can be swapped with alternatives or skipped, allowing for further experimentation.   

\begin{figure}
    \centering
    \includegraphics[width=\textwidth]{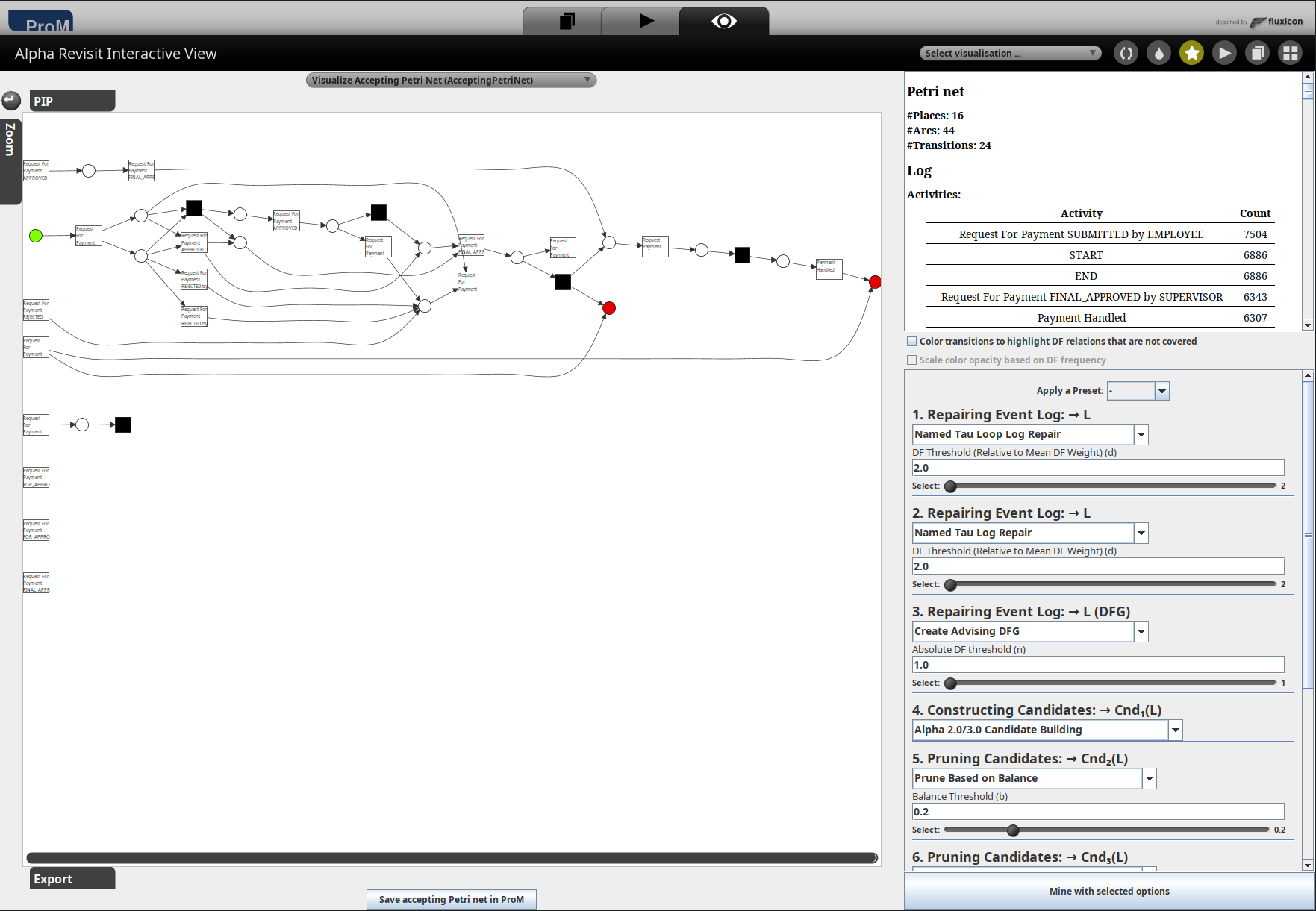}
    \caption{A screenshot of the interactive mode of the developed ProM plugin \textit{Alpha Revisit Experiments}.
    On the right, the steps and different step parameters can be configured.
    The main section on the left shows the discovered Petri net.}
    \label{fig:prom-screenshot}
\end{figure}

\begin{figure}
    \centering
    \includegraphics[width=0.4\textwidth]{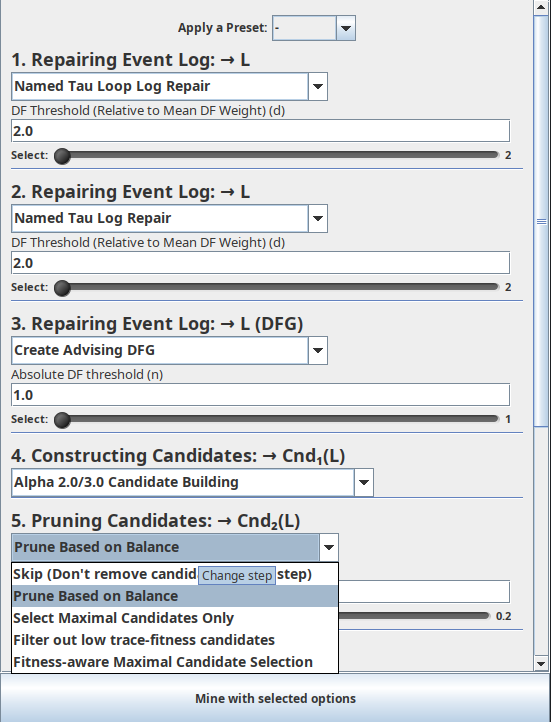}
    \caption{A closeup of the configuration panel of the ProM plugin.
    For each step, there are multiple possible algorithm implementations available, which can be selected from a dropdown-menu.
    Different presets that form a complete process discovery algorithm, e.g., the Alpha+++ algorithm presented here, can be applied at the top.}
    \label{fig:prom-screenshot-options}
\end{figure}

%% file: evaluation.tex
To evaluate the proposed Alpha+++ algorithm ($\alpha$+++), we discovered Petri nets for five real-life event logs, shown in \autoref{tab:evaluation-logs}.
For comparison, we also discovered models using the Inductive Miner Infrequent (IMf) and the standard Alpha algorithm ($\alpha$).
We subsequently calculated alignment-based fitness, precision and F1-scores using PM4Py\footnote{\url{https://pm4py.fit.fraunhofer.de/} (Version 2.6.1)}.

\begin{table}[tbp]
\caption{Overview of the event logs used for evaluation.
        We used a random sample of 3000 cases from the BPI Challenge 2019 log for computational reasons, as it allowed for alignment-based evaluation of the discovered models.
         }
\label{tab:evaluation-logs}
  \centering
  \small
\begin{tabular}{l|c|c|c|c|c}
\textbf{Event Log}                                                                  & \textbf{\#Events} & \textbf{\#Activities} & \textbf{\#Traces} & \textbf{\#Variants} & \textbf{Reference}  \\ \hline
RTFM                                                                                & 561,470           & 11                    & 150,370           & 231                 & \cite{rtfm_log}     \\ \hline
Sepsis                                                                              & 15,214            & 16                    & 1,050             & 846                 & \cite{sepsis_log}   \\ \hline
\begin{tabular}[c]{@{}l@{}}BPI Challenge 2019\\ (Sample of 3000 Cases)\end{tabular} & 18,972            & 34                    & 3,000             & 470                 & \cite{bpic2019_log} \\ \hline
\begin{tabular}[c]{@{}l@{}}BPI Challenge 2020\\ (Request for Payment)\end{tabular}  & 36,796            & 19                    & 6,886             & 89                  & \cite{bpic2020_rfp_log} \\ \hline
\begin{tabular}[c]{@{}l@{}}BPI Challenge 2020\\ (Domestic Declaration)\end{tabular} & 56,437            & 17                    & 10,500            & 99                  & \cite{bpic2020_dd_log}
\end{tabular}
\end{table}

For IMf, we evaluated four models per event log using noise thresholds of $0.1$, $0.2$, $0.3$ and $0.4$.
For $\alpha$, we used four variant filtering approaches upfront: Either only selecting the 10 most common variants or the $n$ most common variants to cover at least $10\%$, $50\%$ or $80\%$ of traces. 
For $\alpha$+++, we chose artificial activity thresholds of $2$ and $4$ (relative to the mean directly-follows weight) for the log repair steps.
Here, a lower threshold value causes more artificial activities to be added.
For each artificial activity threshold, we selected five combinations of the balance $b$, local candidate fitness $t$ and local place replay fitness $r$ thresholds.
Note, that for $t$ and $r$ a value closer to $1$ and for $b$ a value closer to $0$ is more restrictive.
We did not apply problematic activities filtering.

The evaluation results are shown in \autoref{tab:evaluation-results}.
Overall, the fitness and F1-scores of $\alpha$+++ are competitive compared to the IMf.
8 of the 20 models discovered with $\alpha$ are not easy sound (i.e., no final marking is reachable), and thus no alignment scores could be computed.
The remaining 12 models exhibit rather low fitness for some logs but very high precision across the board, significantly boosting the corresponding F1-values. 
Although our approach does not guarantee easy soundness, all 50 Petri nets discovered with $\alpha$+++ are easy sound and allow computation of alignments.
There are notable differences across the different event logs: $\alpha$+++ performs significantly worse compared to the IMf on the Sepsis log in terms of F1-score, caused by lower precision scores, as the models discovered with $\alpha$+++ seem to be underfitting.
On the two BPI Challenge 2020 logs, $\alpha$+++ outperforms the IMf in most configurations, often also exhibiting better fitness and precision scores simultaneously.

The influence of the parameters of $\alpha$+++ is mostly as expected:
More restrictive $b,t,r$ values improve the fitness of the models while decreasing the precision.
\input{evaluation-table.tex}

Manual inspection of the discovered models reveals that the models discovered with $\alpha$+++ are mostly rather simple and often consist of several disconnected model fragments.
Furthermore, multiple models exhibit redundant structures involving silent transitions (e.g., a place with one labeled transition as preset and one silent transition as postset).
Such constructs could be removed by further post-processing of the Petri net.
For more details and a comprehensive list of the discovered models, see \autoref{appendix:disconnected-transitions} and \autoref{appendix:discovered-models} of the appendix.

%% file: evaluation-table.tex
\newcolumntype{X}{>{\centering\let\newline\\\arraybackslash}b{1cm}}
\newcommand\nanVal{\---}

\begin{sidewaystable}[htbp]
 \centering
\caption{Evaluation Results}
    \label{tab:evaluation-results}
\scalebox{0.8}{
\begin{tabular}{l|XXXX|XXXX|XXXXXXXXXX|}
\cline{2-19}
 &
  \multicolumn{4}{c|}{\textbf{Inductive Miner Infrequent}} &
  \multicolumn{4}{c|}{\textbf{Alpha Algorithm}} &
  \multicolumn{10}{c|}{\textbf{Alpha+++ Algorithm}} \\ \cline{2-19} 
 &
  \multicolumn{4}{c|}{Noise Threshold} &
  \multicolumn{4}{c|}{Variant Filtering} &
  \multicolumn{5}{c|}{\textit{Artificial Activity Threshold of 2.0}} &
  \multicolumn{5}{c|}{\textit{Artificial Activity Threshold of 4.0}} \\ \cline{10-19} 
 &
  \multicolumn{1}{c}{0.1} &
  \multicolumn{1}{c}{0.2} &
  \multicolumn{1}{c}{0.3} &
  \multicolumn{1}{c|}{0.4} &
  \multicolumn{1}{c}{Top10} &
  \multicolumn{1}{c}{10\%} &
  \multicolumn{1}{c}{50\%} &
  \multicolumn{1}{c|}{80\%} &
  \multicolumn{1}{c}{\begin{tabular}[c]{@{}c@{}}$b{=}0.5$\\ $t{=}0.5$\\ $r{=}0.5$\end{tabular}} &
  \multicolumn{1}{c}{\begin{tabular}[c]{@{}c@{}}$b{=}0.3$\\ $t{=}0.7$\\ $r{=}0.6$\end{tabular}} &
  \multicolumn{1}{c}{\begin{tabular}[c]{@{}c@{}}$b{=}0.2$\\ $t{=}0.8$\\ $r{=}0.7$\end{tabular}} &
  \multicolumn{1}{c}{\begin{tabular}[c]{@{}c@{}}$b{=}0.2$\\ $t{=}0.8$\\ $r{=}0.8$\end{tabular}} &
  \multicolumn{1}{c|}{\begin{tabular}[c]{@{}c@{}}$b{=}0.1$\\ $t{=}0.9$\\ $r{=}0.9$\end{tabular}} &
  \multicolumn{1}{c}{\begin{tabular}[c]{@{}c@{}}$b{=}0.5$\\ $t{=}0.5$\\ $r{=}0.5$\end{tabular}} &
  \multicolumn{1}{c}{\begin{tabular}[c]{@{}c@{}}$b{=}0.3$\\ $t{=}0.7$\\ $r{=}0.6$\end{tabular}} &
  \multicolumn{1}{c}{\begin{tabular}[c]{@{}c@{}}$b{=}0.2$\\ $t{=}0.8$\\ $r{=}0.7$\end{tabular}} &
  \multicolumn{1}{c}{\begin{tabular}[c]{@{}c@{}}$b{=}0.2$\\ $t{=}0.8$\\ $r{=}0.8$\end{tabular}} &
  \multicolumn{1}{c|}{\begin{tabular}[c]{@{}c@{}}$b{=}0.1$\\ $t{=}0.9$\\ $r{=}0.9$\end{tabular}} \\ \hline
\multicolumn{1}{|l|}{\textbf{RTFM}} &
   &
   &
   &
   &
   &
   &
   &
   &
   &
   &
   &
   &
  \multicolumn{1}{l|}{} &
   &
   &
   &
   &
   \\
\multicolumn{1}{|l|}{Fitness} &
  0.9871 &
  0.9094 &
  0.9091 &
  0.7657 &
  0.6711 &
  0.6731 &
  0.8769 &
  0.8769 &
  0.7880 &
  0.9412 &
  0.9935 &
  0.9935 &
  \multicolumn{1}{l|}{0.9998} &
  0.9160 &
  0.9925 &
  0.9935 &
  0.9935 &
  0.9998 \\
\multicolumn{1}{|l|}{Precision} &
  0.6218 &
  0.6705 &
  0.7959 &
  0.9929 &
  0.6797 &
  1.0000 &
  1.0000 &
  1.0000 &
  0.5223 &
  0.3677 &
  0.3082 &
  0.3082 &
  \multicolumn{1}{l|}{0.3086} &
  0.4056 &
  0.3127 &
  0.3082 &
  0.3082 &
  0.3086 \\
\multicolumn{1}{|l|}{F1 Score} &
  0.7630 &
  0.7719 &
  0.8487 &
  0.8646 &
  0.6754 &
  0.8046 &
  0.9344 &
  0.9344 &
  0.6282 &
  0.5289 &
  0.4705 &
  0.4705 &
  \multicolumn{1}{l|}{0.4716} &
  0.5622 &
  0.4756 &
  0.4705 &
  0.4705 &
  0.4716 \\ \hline
\multicolumn{1}{|l|}{\textbf{Sepsis Cases}} &
   &
   &
   &
   &
   &
   &
   &
   &
   &
   &
   &
   &
  \multicolumn{1}{l|}{} &
   &
   &
   &
   &
   \\
\multicolumn{1}{|l|}{Fitness} &
  0.9382 &
  0.9075 &
  0.8421 &
  0.8108 &
  0.6378 &
  \nanVal &
  \nanVal &
  \nanVal &
  0.9183 &
  0.9362 &
  0.9828 &
  0.9965 &
  \multicolumn{1}{l|}{0.9965} &
  0.9275 &
  0.9636 &
  0.9948 &
  0.9948 &
  1.0000 \\
\multicolumn{1}{|l|}{Precision} &
  0.6049 &
  0.6158 &
  0.6298 &
  0.7285 &
  0.9916 &
  \nanVal &
  \nanVal &
  \nanVal &
  0.3758 &
  0.2922 &
  0.3152 &
  0.2633 &
  \multicolumn{1}{l|}{0.2633} &
  0.2855 &
  0.2923 &
  0.2923 &
  0.2923 &
  0.2805 \\
\multicolumn{1}{|l|}{F1 Score} &
  0.7356 &
  0.7337 &
  0.7206 &
  0.7675 &
  0.7763 &
  \nanVal &
  \nanVal &
  \nanVal &
  0.5334 &
  0.4454 &
  0.4773 &
  0.4166 &
  \multicolumn{1}{l|}{0.4166} &
  0.4365 &
  0.4485 &
  0.4518 &
  0.4518 &
  0.4381 \\ \hline
\multicolumn{1}{|l|}{\textbf{\begin{tabular}[c]{@{}l@{}}BPI Challenge 2019\\ (Sample of 3000 Cases)\end{tabular}}} &
   &
   &
   &
   &
   &
   &
   &
   &
   &
   &
   &
   &
  \multicolumn{1}{l|}{} &
   &
   &
   &
   &
   \\
\multicolumn{1}{|l|}{Fitness} &
  0.9906 &
  0.9938 &
  0.9536 &
  0.9177 &
  0.6282 &
  0.7462 &
  \nanVal &
  0.3205 &
  0.9422 &
  0.9431 &
  0.9506 &
  0.9506 &
  \multicolumn{1}{l|}{1.0000} &
  0.9422 &
  0.9433 &
  0.9602 &
  0.9602 &
  1.0000 \\
\multicolumn{1}{|l|}{Precision} &
  0.2086 &
  0.2379 &
  0.2383 &
  0.2528 &
  0.9986 &
  1.0000 &
  \nanVal &
  0.9964 &
  0.3395 &
  0.3180 &
  0.2416 &
  0.2416 &
  \multicolumn{1}{l|}{0.1968} &
  0.3748 &
  0.3487 &
  0.2501 &
  0.2501 &
  0.1968 \\
\multicolumn{1}{|l|}{F1 Score} &
  0.3446 &
  0.3839 &
  0.3813 &
  0.3964 &
  0.7712 &
  0.8547 &
  \nanVal &
  0.4850 &
  0.4992 &
  0.4757 &
  0.3852 &
  0.3852 &
  \multicolumn{1}{l|}{0.3288} &
  0.5363 &
  0.5092 &
  0.3968 &
  0.3968 &
  0.3288 \\ \hline
\multicolumn{1}{|l|}{\textbf{\begin{tabular}[c]{@{}l@{}}BPI Challenge 2020\\ (Requests for Payment)\end{tabular}}} &
   &
   &
   &
   &
   &
   &
   &
   &
   &
   &
   &
   &
  \multicolumn{1}{l|}{} &
   &
   &
   &
   &
   \\
\multicolumn{1}{|l|}{Fitness} &
  0.9476 &
  0.9051 &
  0.9051 &
  0.9051 &
  \nanVal &
  0.8678 &
  0.8380 &
  \nanVal &
  0.9179 &
  0.9438 &
  0.9438 &
  0.9438 &
  \multicolumn{1}{l|}{0.9595} &
  0.9179 &
  0.9438 &
  0.9438 &
  0.9438 &
  0.9595 \\
\multicolumn{1}{|l|}{Precision} &
  0.3173 &
  0.2704 &
  0.2704 &
  0.2704 &
  \nanVal &
  1.0000 &
  1.0000 &
  \nanVal &
  0.5415 &
  0.4451 &
  0.4451 &
  0.4451 &
  \multicolumn{1}{l|}{0.3500} &
  0.5415 &
  0.4451 &
  0.4451 &
  0.4451 &
  0.3500 \\
\multicolumn{1}{|l|}{F1 Score} &
  0.4754 &
  0.4164 &
  0.4164 &
  0.4164 &
  \nanVal &
  0.9292 &
  0.9119 &
  \nanVal &
  0.6812 &
  0.6049 &
  0.6049 &
  0.6049 &
  \multicolumn{1}{l|}{0.5129} &
  0.6812 &
  0.6049 &
  0.6049 &
  0.6049 &
  0.5129 \\ \hline
\multicolumn{1}{|l|}{\textbf{\begin{tabular}[c]{@{}l@{}}BPI Challenge 2020\\ (Domestic Declaration)\end{tabular}}} &
   &
   &
   &
   &
   &
   &
   &
   &
   &
   &
   &
   &
  \multicolumn{1}{l|}{} &
   &
   &
   &
   &
   \\
\multicolumn{1}{|l|}{Fitness} &
  0.9499 &
  0.9302 &
  0.9302 &
  0.9302 &
  \nanVal &
  0.8906 &
  0.8549 &
  \nanVal &
  0.9029 &
  0.9265 &
  0.9308 &
  0.9308 &
  \multicolumn{1}{l|}{0.9493} &
  0.9143 &
  0.9143 &
  0.9461 &
  0.9461 &
  0.9477 \\
\multicolumn{1}{|l|}{Precision} &
  0.4056 &
  0.2469 &
  0.2469 &
  0.2469 &
  \nanVal &
  1.0000 &
  1.0000 &
  \nanVal &
  0.9094 &
  0.7206 &
  0.7192 &
  0.7192 &
  \multicolumn{1}{l|}{0.4897} &
  0.6795 &
  0.6795 &
  0.4780 &
  0.4780 &
  0.4780 \\
\multicolumn{1}{|l|}{F1 Score} &
  0.5685 &
  0.3902 &
  0.3902 &
  0.3902 &
  \nanVal &
  0.9421 &
  0.9218 &
  \nanVal &
  0.9061 &
  0.8107 &
  0.8114 &
  0.8114 &
  \multicolumn{1}{l|}{0.6461} &
  0.7796 &
  0.7796 &
  0.6351 &
  0.6351 &
  0.6354 \\ \hline
\end{tabular}
}
\end{sidewaystable}

%% file: concl.tex
In this paper, we revisited the Alpha algorithm to overcome its limitations, focusing on real-life event logs.
For that, we presented the Alpha+++ algorithm which, like the Alpha algorithm, primarily uses directly-follows relations to discover Petri nets.
Alpha+++ pre-processes event logs by adding artificial activities for potential loop or skip constructs.
This allows discovering silent transitions while still assessing the fitness of places by easily computable token-based replay instead of expensive alignment computations.
Subsequently, place candidates are generated based on a pruned DFG.
A multistep candidate filtering approach efficiently removes place candidates with low fitness, configurable through parameters.
We implemented the Alpha+++ algorithm both as a ProM plugin and in Python.
The ProM plugin is available in ProM nightly builds and also features an interactive mode to allow experimenting with different algorithm steps and parameters.
We evaluated the Alpha+++ on five real-life event logs and compared the results to the classical Alpha algorithm and the widely adapted Inductive Miner Infrequent.
Overall, the results indicate that the Alpha+++ algorithm is competitive in terms of fitness and precision.
In general, the different step parameter configurations tested reliably determine the trade-off between fitness and precision.

Further research should include further evaluation of the algorithm.
For that, additional performance metrics like simplicity or generality could be included and also compared to other process discovery algorithms.
It is particularly interesting to see if there are any patterns regarding algorithm parameters, event log properties, and model performance.
Such observations could enable automatic parameter selection based on the log, and thus simplify Alpha+++ to a well-performing one-in-all algorithm.
Additionally, a more comprehensive qualitative analysis of the discovered models is needed.
Further research could also explore if any theoretical guarantees, such as easy-soundness, are attainable, e.g., using more sophisticated post-processing of the discovered Petri net.

%% file: appendix.tex
\clearpage

\section{Generality and Simplicity Evaluation}
\input{appendix-simplicity-generalization-table.tex}

\clearpage

\section{Disconnected Transitions}
\label{appendix:disconnected-transitions}
In many of the models discovered with Alpha+++ there are one or more transitions that are disconnected from the rest of the model.
As expected, this phenomenon is most frequently observable for more restricting algorithm parameters, which aggressively filter out low-fitness place candidates and places.
In \autoref{app:tab:disconnected-trans}, we counted the number of disconnected labeled transitions for each of the models and calculated what percentage of the activities of the log are represented by disconnected transitions on average. 

\newcolumntype{Y}{>{\centering\let\newline\\\arraybackslash}p{1cm}}
\begin{table}[]
  \centering
  \caption{Disconnected labeled transitions in models discovered with Alpha+++. The average percentage refers to the average fraction of activities in the event log that are labels of disconnected transitions.}
  \label{app:tab:disconnected-trans}
  \scalebox{0.8}{
\begin{tabular}{l|YYYYYYYYYY|c|lllllll}
\cline{2-12}
 &
  \multicolumn{10}{c|}{\textbf{Alpha+++ Algorithm}} &
  \multirow{3}{*}{\textbf{Average \%}} &
  \multicolumn{1}{c}{} &
  \multicolumn{1}{c}{} &
  \multicolumn{1}{c}{} &
  \multicolumn{1}{c}{} &
  \multicolumn{1}{c}{} &
  \multicolumn{1}{c}{} &
  \multicolumn{1}{c}{} \\ \cline{2-11}
 &
  \multicolumn{5}{c|}{\textit{Artificial Activity Threshold of 2.0}} &
  \multicolumn{5}{c|}{\textit{Artificial Activity Threshold of 4.0}} &
   &
  \multicolumn{1}{c}{\textit{}} &
  \multicolumn{1}{c}{\textit{}} &
  \multicolumn{1}{c}{} &
  \multicolumn{1}{c}{} &
  \multicolumn{1}{c}{} &
  \multicolumn{1}{c}{} &
   \\
 &
  \begin{tabular}[c]{@{}c@{}}$b{=}0.5$\\ $t{=}0.5$\\ $r{=}0.5$\end{tabular} &
  \begin{tabular}[c]{@{}c@{}}$b{=}0.3$\\ $t{=}0.7$\\ $r{=}0.6$\end{tabular} &
  \begin{tabular}[c]{@{}c@{}}$b{=}0.2$\\ $t{=}0.8$\\ $r{=}0.7$\end{tabular} &
  \begin{tabular}[c]{@{}c@{}}$b{=}0.2$\\ $t{=}0.8$\\ $r{=}0.8$\end{tabular} &
  \multicolumn{1}{c|}{\begin{tabular}[c]{@{}c@{}}$b{=}0.1$\\ $t{=}0.9$\\ $r{=}0.9$\end{tabular}} &
  \begin{tabular}[c]{@{}c@{}}$b{=}0.5$\\ $t{=}0.5$\\ $r{=}0.5$\end{tabular} &
  \begin{tabular}[c]{@{}c@{}}$b{=}0.3$\\ $t{=}0.7$\\ $r{=}0.6$\end{tabular} &
  \begin{tabular}[c]{@{}c@{}}$b{=}0.2$\\ $t{=}0.8$\\ $r{=}0.7$\end{tabular} &
  \begin{tabular}[c]{@{}c@{}}$b{=}0.2$\\ $t{=}0.8$\\ $r{=}0.8$\end{tabular} &
  \begin{tabular}[c]{@{}c@{}}$b{=}0.1$\\ $t{=}0.9$\\ $r{=}0.9$\end{tabular} &
   &
   &
   &
   &
   &
   &
   &
   \\ \cline{1-12}
\multicolumn{1}{|l|}{RTFM} &
  1 &
  1 &
  2 &
  2 &
  \multicolumn{1}{c|}{2} &
  1 &
  1 &
  2 &
  2 &
  2 &
  14.55 &
   &
   &
   &
   &
   &
   &
   \\ \cline{1-12}
\multicolumn{1}{|l|}{Sepsis Cases} &
  1 &
  5 &
  5 &
  7 &
  \multicolumn{1}{c|}{7} &
  6 &
  7 &
  12 &
  12 &
  13 &
  46.88 &
   &
   &
   &
   &
   &
   &
   \\ \cline{1-12}
\multicolumn{1}{|l|}{\begin{tabular}[c]{@{}l@{}}BPI Challenge 2019\\ (Sample of 3000 Cases)\end{tabular}} &
  8 &
  10 &
  14 &
  14 &
  \multicolumn{1}{c|}{22} &
  7 &
  10 &
  14 &
  14 &
  24 &
  40.29 &
   &
   &
   &
   &
   &
   &
   \\ \cline{1-12}
\multicolumn{1}{|l|}{\begin{tabular}[c]{@{}l@{}}BPI Challenge 2020\\ (Requests for Payment)\end{tabular}} &
  0 &
  0 &
  0 &
  0 &
  \multicolumn{1}{c|}{0} &
  0 &
  0 &
  0 &
  0 &
  0 &
  0 &
   &
   &
   &
   &
   &
   &
   \\ \cline{1-12}
\multicolumn{1}{|l|}{BPIC 2021} &
  0 &
  0 &
  0 &
  0 &
  \multicolumn{1}{c|}{0} &
  1 &
  1 &
  1 &
  1 &
  1 &
  2.94 &
   &
   &
   &
   &
   &
   &
   \\ \cline{1-12}
\end{tabular}
}
  \end{table}
  
In Alpha+++ disconnected transitions are not handled separately and always included in the final Petri net.
Note, that always including the transitions without restrictions is not the only option, and has non-negligible effects on the fitness and precision of the complete model.
In particular, the decision whether an activity should be included as a disconnected transition, i.e., without restrictions, or not at all can be made separately for each transition based on frequency information of the corresponding activity.

\begin{figure}
    \centering
    \includegraphics[width=\textwidth]{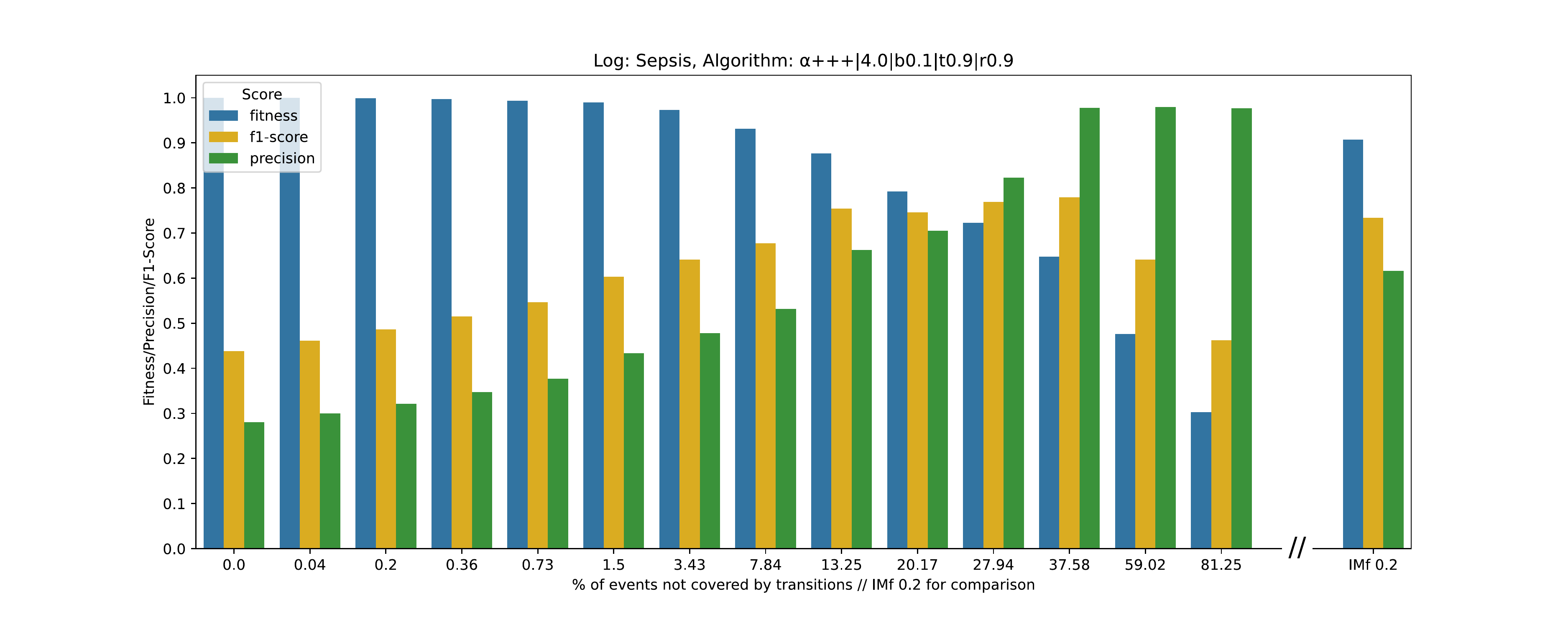}
    \caption{Fitness, precision and F1 score values of the resulting models when removing transitions one by one greedily.
    Clearly, in most steps, the greedy ordering has the desired effect of only decreasing the fitness slightly while increasing the precision significantly. 
    Performance values for the Inductive Miner Infrequent 0.2 are given as a reference point.}
    \label{fig:discon_acts_Sepsis}
\end{figure}

In particular, certain disconnected transitions can be omitted with the goal to maximize certain performance metrics, like precision or F1 score.
Note that the fitness of the model will consistently decrease whenever a disconnected (labeled) transition is removed, as the resulting Petri net is more restrictive, i.e., not allowing this activity to occur at all.
Thus, it is in particular interesting to attempt to increase the precision of the end model while not decreasing the fitness too much. 
As a heuristic, we propose removing the labeled, disconnected transitions in the reverse order of their frequency in the log (i.e., transitions with activity labels that are infrequent in the log are removed first).
We call this technique of determining the removal order \emph{greedy}.
Removing a transition with a rather infrequent activity label often leads to only a small decrease in fitness and a bigger gain in precision.
In \autoref{fig:discon_acts_Sepsis} we applied this technique to a model discovered using the Sepsis event log.
As such, we removed disconnected, labeled transitions one by one in the given order, and measured the fitness, precision and F1 score in every step.
The x-axis additionally shows the number of events that are no longer covered by a transition (i.e., the sum of events in the log with an activity label corresponding to a transition that was removed).

We ran experiments to compare how well the greedy activity removal performs compared to the brute-force approach of evaluating all possible removal orders.
For that, we selected to maximize the F1 score and evaluated the Petri nets obtained by removing the transitions corresponding to the first $k$ infrequent activities.
In the brute force approach, we also selected a set of $k$ disconnected labeled transitions to remove, but choose that set from all possibilities that maximizes the F1 score of the resulting net.
This allows us to compare the F1 score for every number $k$ of removed transitions between the removal set based on the greedy ordering or the best possibility for maximizing the F1 score.
As such, the brute force approach provides an upper bound of achievable F1 score improvements and is always guaranteed to be at least as high as the F1 score achieved using greedy removal.

\begin{figure}[h]
    \centering
    \includegraphics[width=1.2\textwidth]{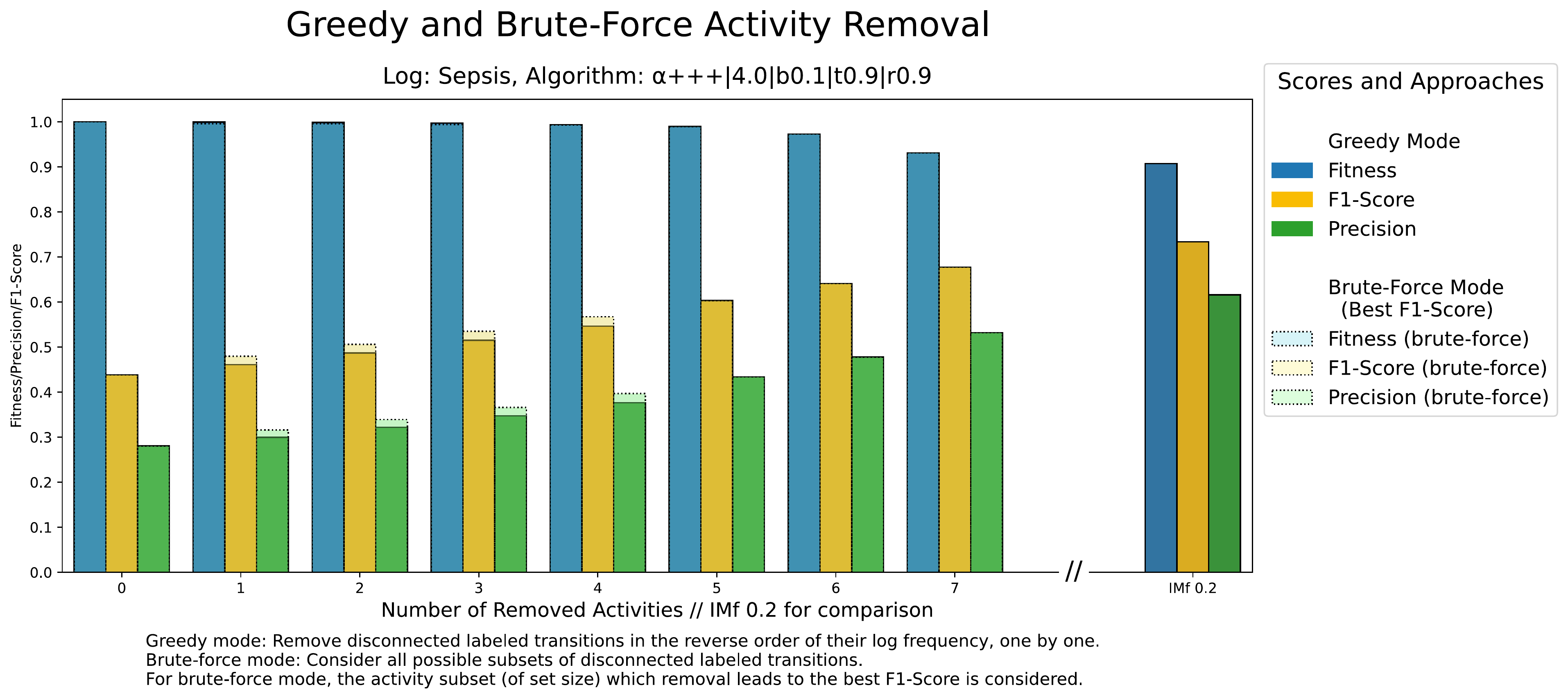}
    \caption{Comparison of the greedy activity removal order versus the best removal set of that size (w.r.t. F1 score).
    The possible F1 score improvements of the brute-force approach (shown as light orange on top of bar) are noticeable for some removal counts.
    } 
    \label{fig:discon_acts_Sepsis_comparison}
\end{figure}

\autoref{fig:discon_acts_Sepsis_comparison} shows the observed gain in F1 score from the brute-force approach compared to the greedy ordering for one model discovered with the Sepsis event log.
Our experiments indicate that for most use-cases, the greedy activity removal order should be preferred because of the significantly better running time.
While the brute-force approach might find better removal sets in some configurations, it requires evaluating models for every possible removal set of a given size, which is computationally expensive.
In particular, we observed that for many discovered models the greedy removal order was, in fact, already optimal regarding the F1 score.

The brute-force and greedy transition removal post-processing, as well as the corresponding evaluation and visualization functions, are also published open-source\footnote{\url{https://github.com/aarkue/alpha-revisit-python/blob/main/remove-disconnected-acts.ipynb}}.

\clearpage

\section{Discovered Process Models}
\label{appendix:discovered-models}
\newcommand\appendixFigure[3]{
  \begin{figure*}[h!]
    \centering
    \includegraphics[width=1.2\textwidth,height=0.5\textwidth,keepaspectratio]{./figures/appendix/#2}
    \caption{#1}
  \end{figure*}
}

In the next pages, we include all the process models used for evaluation (see \autoref{sec:eval}).
The logs are enumerated in the following order:
\begin{enumerate}
  \item \nameref{appendix:log:RTFM}
  \item \nameref{appendix:log:Sepsis}
  \item \nameref{appendix:log:BPI_Challenge_2019_sampled_3000cases}
  \item \nameref{appendix:log:BPI_Challenge_2020}
  \item \nameref{appendix:log:BPI_Challenge_2020_DomesticDeclarations}
\end{enumerate}

To stay concise, we abbreviate the algorithms as before and additionally introduce the following notation for the different configurations of the Alpha+++ algorithm:
\begin{align*}
  \alpha\text{+++};\underbrace{2.0}_\text{DF Threshold};b\underbrace{0.3}_\text{Balance Thresh.};t\underbrace{0.7}_\text{Local Candidate Fitness Thresh.};r\underbrace{0.6}_\text{Replay Fitness Thresh.}
\end{align*}
For example, the configuration \texttt{$\alpha{}$+++;2.0;b0.3;t0.7;r0.6} represents that the model was discovered with $\alpha$+++ using a relative DF threshold of 2.0 (used for adding artificial activities), a balance threshold of $0.3$ (used to filter place candidates), a local candidate fitness threshold of $0.7$ (used to prune the candidates further) and finally a replay place fitness threshold of $0.6$ (used to post-process the Petri net by removing unfit places).

We (visually) post-processed the models discovered with $\alpha$+++, to allow for easier interpretation and comparison of the models, as the models often exhibit multiple disconnect parts.
In particular, we added artificial start and end transitions, marked by $\actstart$ or $\actend$, and connected them to an added place, which also connects all disconnected process parts.
Note, that this does not alter the underlying semantic of the Petri net, if interpreted correctly.
\subsection{RTFM}
\label{appendix:log:RTFM}
\appendixFigure{Model discovered using \texttt{IMf 0.1} on \textit{RTFM}}{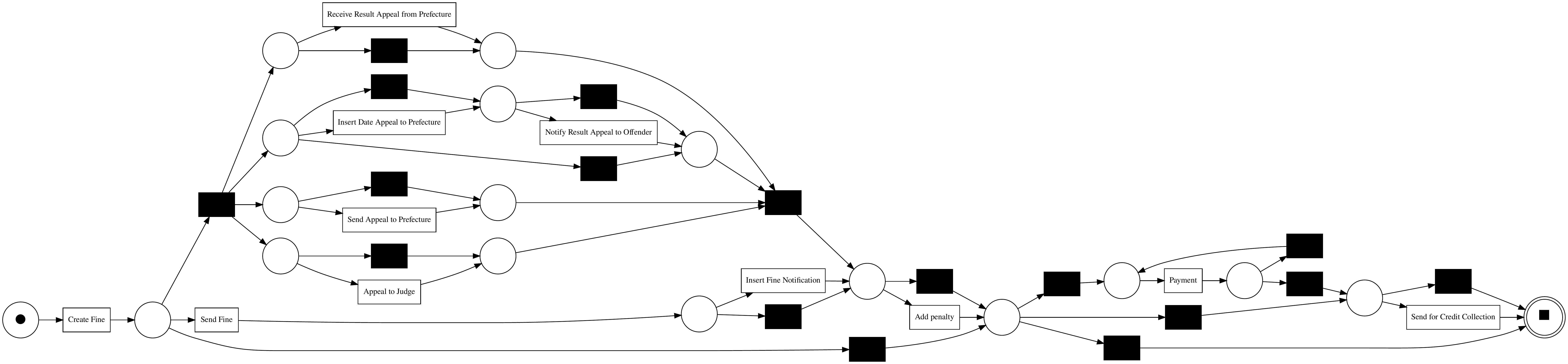}{appendix:fig:RTFM:IMf 0.1}

\appendixFigure{Model discovered using \texttt{IMf 0.2} on \textit{RTFM}}{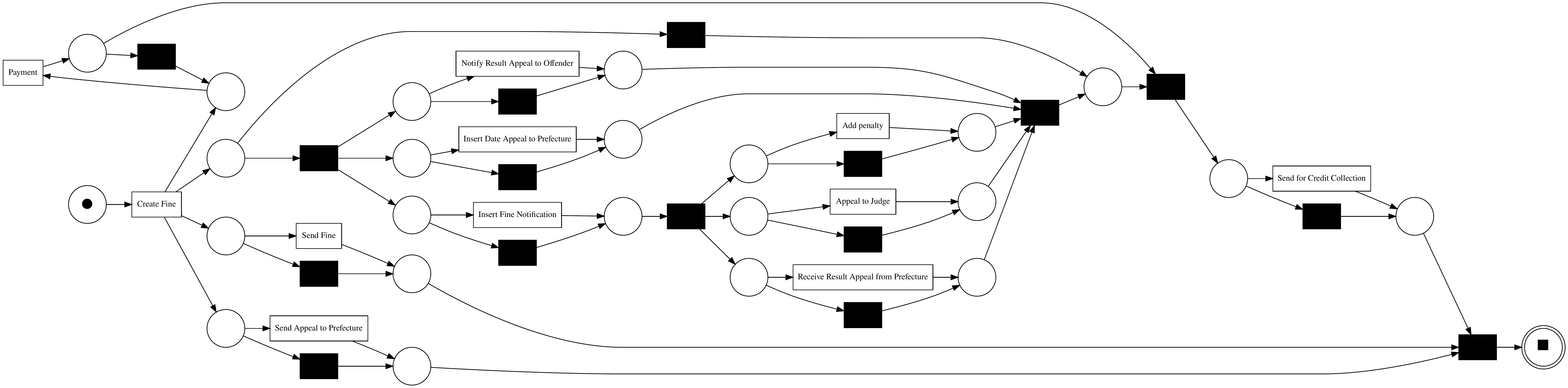}{appendix:fig:RTFM:IMf 0.2}

\appendixFigure{Model discovered using \texttt{IMf 0.3} on \textit{RTFM}}{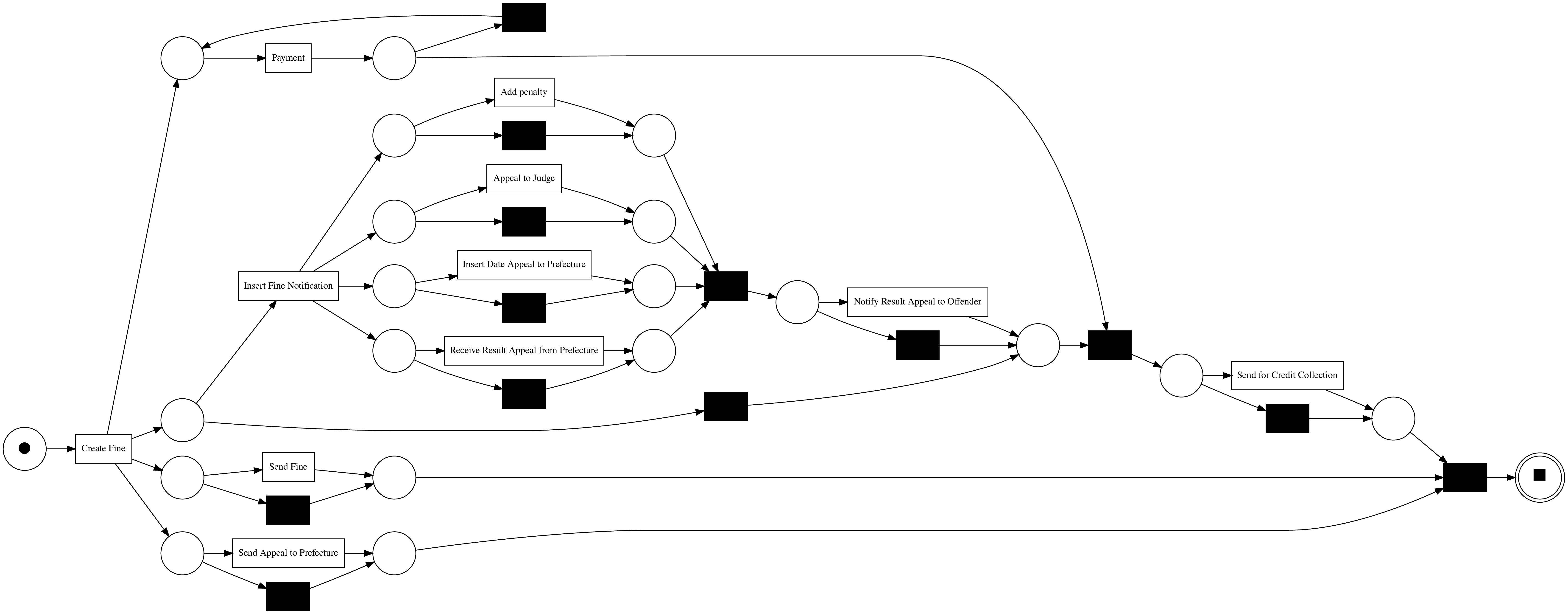}{appendix:fig:RTFM:IMf 0.3}

\appendixFigure{Model discovered using \texttt{IMf 0.4} on \textit{RTFM}}{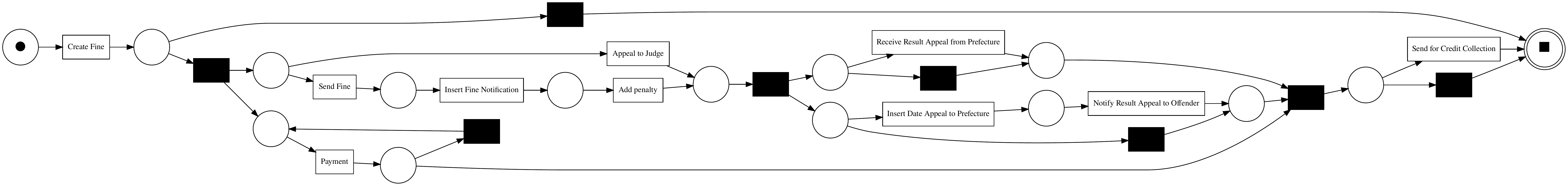}{appendix:fig:RTFM:IMf 0.4}

\appendixFigure{Model discovered using \texttt{$\alpha{}$ Top10} on \textit{RTFM}}{RTFM_model_alpha_Top10.pdf}{appendix:fig:RTFM:α Top10}

\appendixFigure{Model discovered using \texttt{$\alpha{}$ 10\%Cov} on \textit{RTFM}}{RTFM_model_alpha_10PERCENTCov.pdf}{appendix:fig:RTFM:α 10\%Cov}

\appendixFigure{Model discovered using \texttt{$\alpha{}$ 50\%Cov} on \textit{RTFM}}{RTFM_model_alpha_50PERCENTCov.pdf}{appendix:fig:RTFM:α 50\%Cov}

\appendixFigure{Model discovered using \texttt{$\alpha{}$ 80\%Cov} on \textit{RTFM}}{RTFM_model_alpha_80PERCENTCov.pdf}{appendix:fig:RTFM:α 80\%Cov}

\appendixFigure{Model discovered using \texttt{$\alpha{}$+++;2.0;b0.5;t0.5;r0.5} on \textit{RTFM}}{RTFM_model_alphappp-2.0-b0.5-t0.5-r0.5.pdf}{appendix:fig:RTFM:α+++|2.0|b0.5|t0.5|r0.5}

\appendixFigure{Model discovered using \texttt{$\alpha{}$+++;2.0;b0.3;t0.7;r0.6} on \textit{RTFM}}{RTFM_model_alphappp-2.0-b0.3-t0.7-r0.6.pdf}{appendix:fig:RTFM:α+++|2.0|b0.3|t0.7|r0.6}

\appendixFigure{Model discovered using \texttt{$\alpha{}$+++;2.0;b0.2;t0.8;r0.7} on \textit{RTFM}}{RTFM_model_alphappp-2.0-b0.2-t0.8-r0.7.pdf}{appendix:fig:RTFM:α+++|2.0|b0.2|t0.8|r0.7}

\appendixFigure{Model discovered using \texttt{$\alpha{}$+++;2.0;b0.2;t0.8;r0.8} on \textit{RTFM}}{RTFM_model_alphappp-2.0-b0.2-t0.8-r0.8.pdf}{appendix:fig:RTFM:α+++|2.0|b0.2|t0.8|r0.8}

\appendixFigure{Model discovered using \texttt{$\alpha{}$+++;2.0;b0.1;t0.9;r0.9} on \textit{RTFM}}{RTFM_model_alphappp-2.0-b0.1-t0.9-r0.9.pdf}{appendix:fig:RTFM:α+++|2.0|b0.1|t0.9|r0.9}

\appendixFigure{Model discovered using \texttt{$\alpha{}$+++;4.0;b0.5;t0.5;r0.5} on \textit{RTFM}}{RTFM_model_alphappp-4.0-b0.5-t0.5-r0.5.pdf}{appendix:fig:RTFM:α+++|4.0|b0.5|t0.5|r0.5}

\appendixFigure{Model discovered using \texttt{$\alpha{}$+++;4.0;b0.3;t0.7;r0.6} on \textit{RTFM}}{RTFM_model_alphappp-4.0-b0.3-t0.7-r0.6.pdf}{appendix:fig:RTFM:α+++|4.0|b0.3|t0.7|r0.6}

\appendixFigure{Model discovered using \texttt{$\alpha{}$+++;4.0;b0.2;t0.8;r0.7} on \textit{RTFM}}{RTFM_model_alphappp-4.0-b0.2-t0.8-r0.7.pdf}{appendix:fig:RTFM:α+++|4.0|b0.2|t0.8|r0.7}

\appendixFigure{Model discovered using \texttt{$\alpha{}$+++;4.0;b0.2;t0.8;r0.8} on \textit{RTFM}}{RTFM_model_alphappp-4.0-b0.2-t0.8-r0.8.pdf}{appendix:fig:RTFM:α+++|4.0|b0.2|t0.8|r0.8}

\appendixFigure{Model discovered using \texttt{$\alpha{}$+++;4.0;b0.1;t0.9;r0.9} on \textit{RTFM}}{RTFM_model_alphappp-4.0-b0.1-t0.9-r0.9.pdf}{appendix:fig:RTFM:α+++|4.0|b0.1|t0.9|r0.9}

\clearpage
\subsection{Sepsis Cases}
\label{appendix:log:Sepsis}
\appendixFigure{Model discovered using \texttt{IMf 0.1} on \textit{Sepsis Cases}}{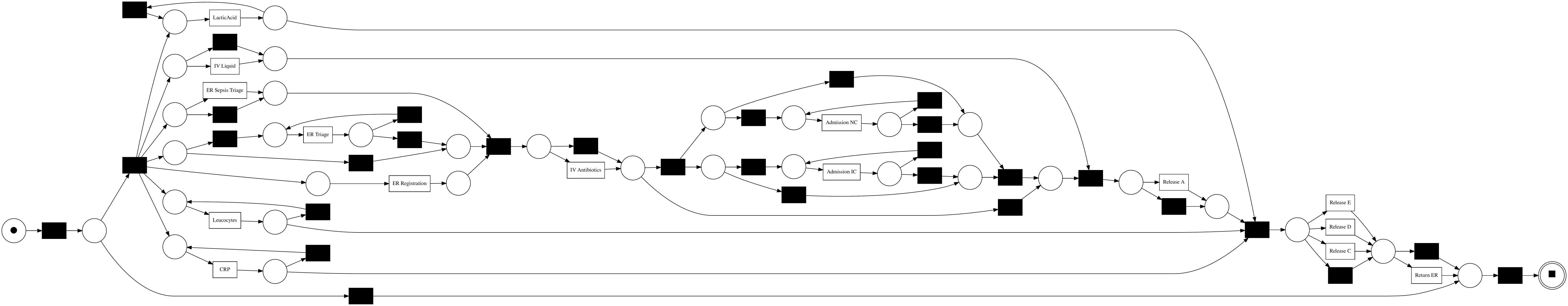}{appendix:fig:Sepsis:IMf 0.1}

\appendixFigure{Model discovered using \texttt{IMf 0.2} on \textit{Sepsis Cases}}{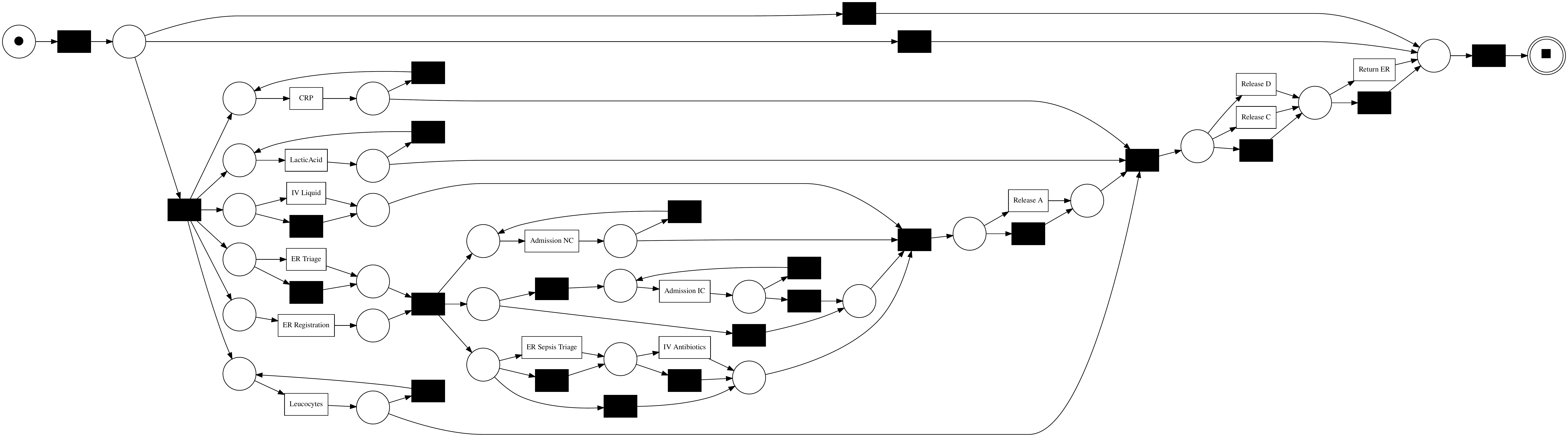}{appendix:fig:Sepsis:IMf 0.2}

\appendixFigure{Model discovered using \texttt{IMf 0.3} on \textit{Sepsis Cases}}{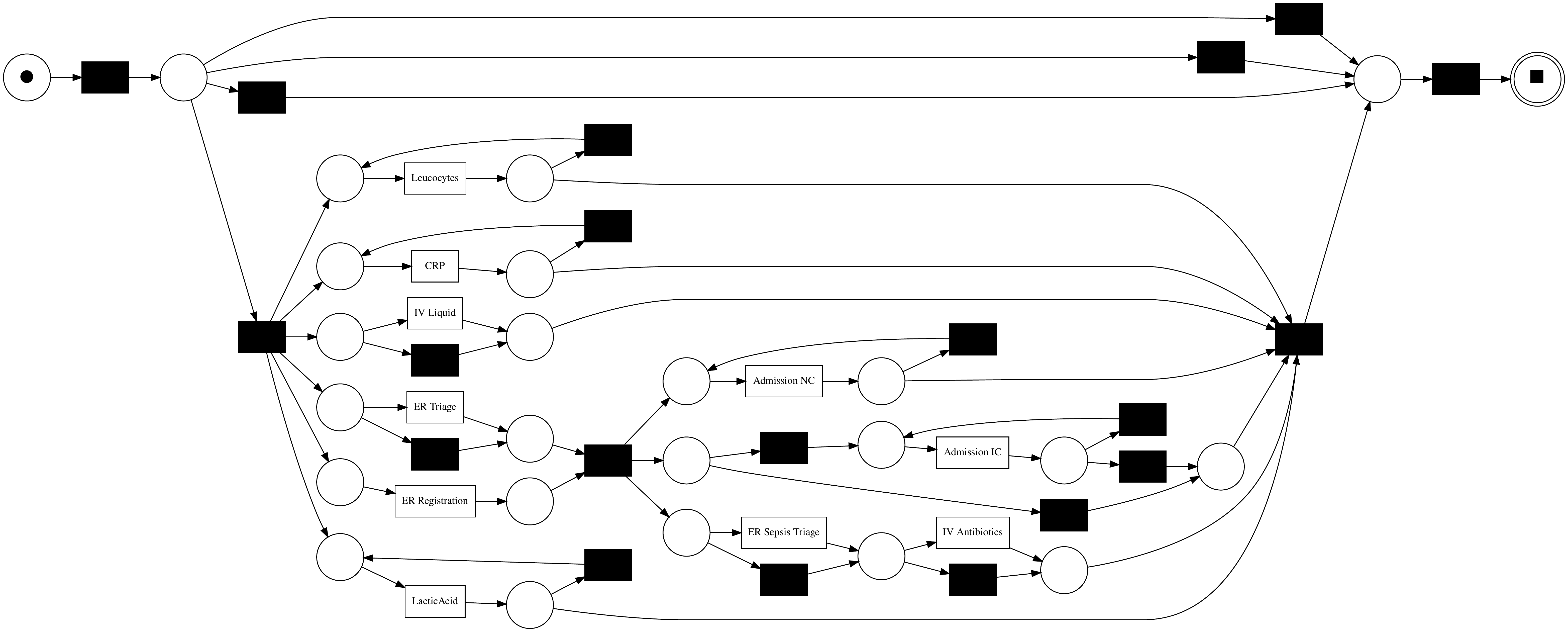}{appendix:fig:Sepsis:IMf 0.3}

\appendixFigure{Model discovered using \texttt{IMf 0.4} on \textit{Sepsis Cases}}{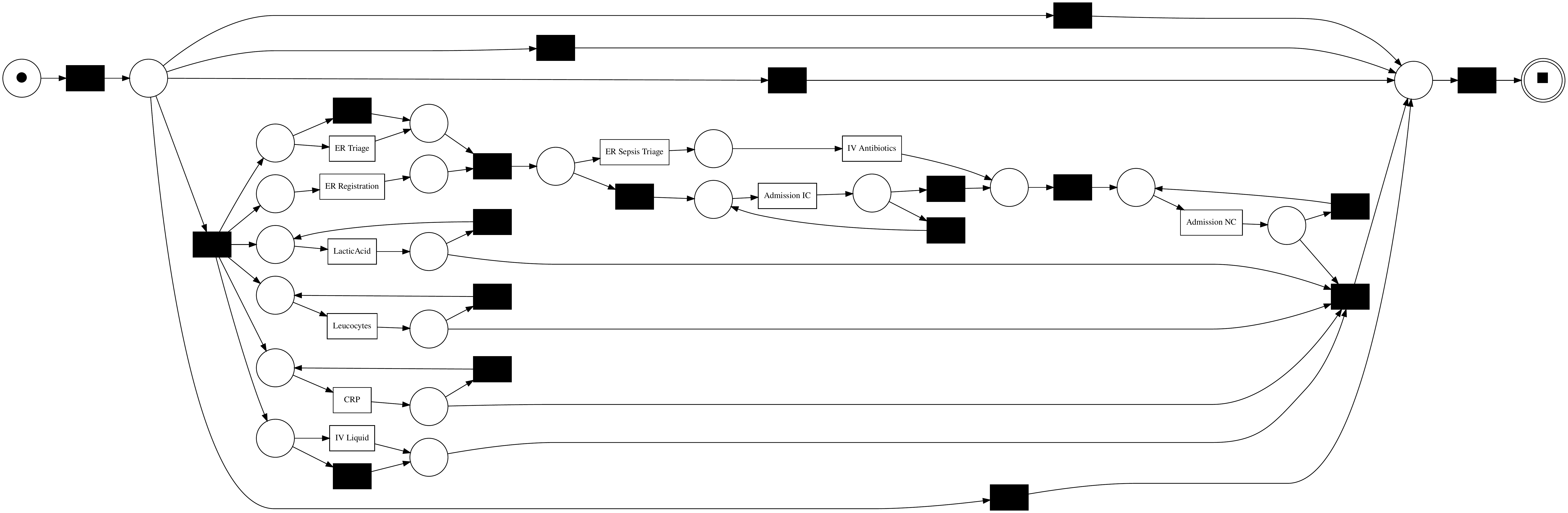}{appendix:fig:Sepsis:IMf 0.4}

\appendixFigure{Model discovered using \texttt{$\alpha{}$ Top10} on \textit{Sepsis Cases}}{Sepsis_model_alpha_Top10.pdf}{appendix:fig:Sepsis:α Top10}

\appendixFigure{Model discovered using \texttt{$\alpha{}$ 10\%Cov} on \textit{Sepsis Cases}}{Sepsis_model_alpha_10PERCENTCov.pdf}{appendix:fig:Sepsis:α 10\%Cov}

\appendixFigure{Model discovered using \texttt{$\alpha{}$ 50\%Cov} on \textit{Sepsis Cases}}{Sepsis_model_alpha_50PERCENTCov.pdf}{appendix:fig:Sepsis:α 50\%Cov}

\appendixFigure{Model discovered using \texttt{$\alpha{}$ 80\%Cov} on \textit{Sepsis Cases}}{Sepsis_model_alpha_80PERCENTCov.pdf}{appendix:fig:Sepsis:α 80\%Cov}

\appendixFigure{Model discovered using \texttt{$\alpha{}$+++;2.0;b0.5;t0.5;r0.5} on \textit{Sepsis Cases}}{Sepsis_model_alphappp-2.0-b0.5-t0.5-r0.5.pdf}{appendix:fig:Sepsis:α+++|2.0|b0.5|t0.5|r0.5}

\appendixFigure{Model discovered using \texttt{$\alpha{}$+++;2.0;b0.3;t0.7;r0.6} on \textit{Sepsis Cases}}{Sepsis_model_alphappp-2.0-b0.3-t0.7-r0.6.pdf}{appendix:fig:Sepsis:α+++|2.0|b0.3|t0.7|r0.6}

\appendixFigure{Model discovered using \texttt{$\alpha{}$+++;2.0;b0.2;t0.8;r0.7} on \textit{Sepsis Cases}}{Sepsis_model_alphappp-2.0-b0.2-t0.8-r0.7.pdf}{appendix:fig:Sepsis:α+++|2.0|b0.2|t0.8|r0.7}

\appendixFigure{Model discovered using \texttt{$\alpha{}$+++;2.0;b0.2;t0.8;r0.8} on \textit{Sepsis Cases}}{Sepsis_model_alphappp-2.0-b0.2-t0.8-r0.8.pdf}{appendix:fig:Sepsis:α+++|2.0|b0.2|t0.8|r0.8}

\appendixFigure{Model discovered using \texttt{$\alpha{}$+++;2.0;b0.1;t0.9;r0.9} on \textit{Sepsis Cases}}{Sepsis_model_alphappp-2.0-b0.1-t0.9-r0.9.pdf}{appendix:fig:Sepsis:α+++|2.0|b0.1|t0.9|r0.9}

\appendixFigure{Model discovered using \texttt{$\alpha{}$+++;4.0;b0.5;t0.5;r0.5} on \textit{Sepsis Cases}}{Sepsis_model_alphappp-4.0-b0.5-t0.5-r0.5.pdf}{appendix:fig:Sepsis:α+++|4.0|b0.5|t0.5|r0.5}

\appendixFigure{Model discovered using \texttt{$\alpha{}$+++;4.0;b0.3;t0.7;r0.6} on \textit{Sepsis Cases}}{Sepsis_model_alphappp-4.0-b0.3-t0.7-r0.6.pdf}{appendix:fig:Sepsis:α+++|4.0|b0.3|t0.7|r0.6}

\appendixFigure{Model discovered using \texttt{$\alpha{}$+++;4.0;b0.2;t0.8;r0.7} on \textit{Sepsis Cases}}{Sepsis_model_alphappp-4.0-b0.2-t0.8-r0.7.pdf}{appendix:fig:Sepsis:α+++|4.0|b0.2|t0.8|r0.7}

\appendixFigure{Model discovered using \texttt{$\alpha{}$+++;4.0;b0.2;t0.8;r0.8} on \textit{Sepsis Cases}}{Sepsis_model_alphappp-4.0-b0.2-t0.8-r0.8.pdf}{appendix:fig:Sepsis:α+++|4.0|b0.2|t0.8|r0.8}

\appendixFigure{Model discovered using \texttt{$\alpha{}$+++;4.0;b0.1;t0.9;r0.9} on \textit{Sepsis Cases}}{Sepsis_model_alphappp-4.0-b0.1-t0.9-r0.9.pdf}{appendix:fig:Sepsis:α+++|4.0|b0.1|t0.9|r0.9}

\clearpage
\subsection{BPI Challenge 2019 (Sample of 3000 Cases)}
\label{appendix:log:BPI_Challenge_2019_sampled_3000cases}
\appendixFigure{Model discovered using \texttt{IMf 0.1} on \textit{BPI Challenge 2019 (Sample of 3000 Cases)}}{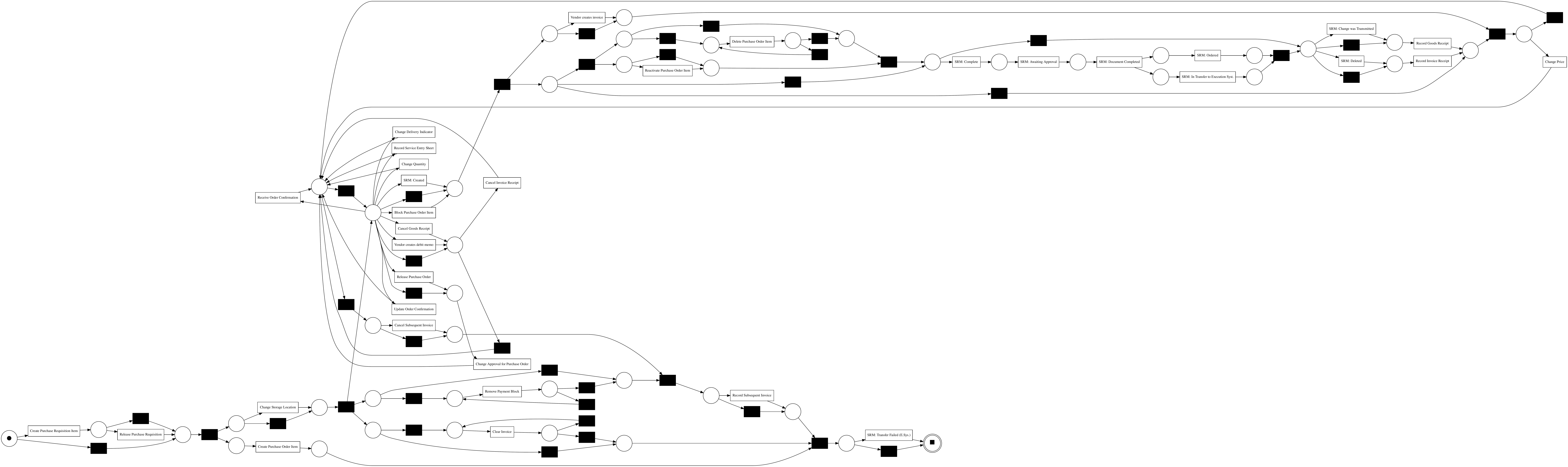}{appendix:fig:BPI_Challenge_2019_sampled_3000cases:IMf 0.1}

\appendixFigure{Model discovered using \texttt{IMf 0.2} on \textit{BPI Challenge 2019 (Sample of 3000 Cases)}}{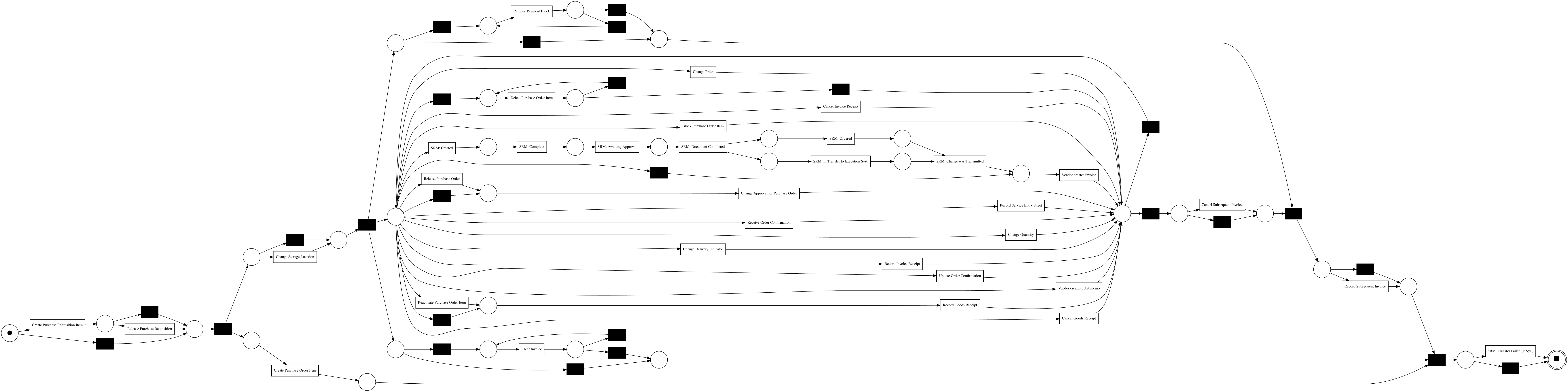}{appendix:fig:BPI_Challenge_2019_sampled_3000cases:IMf 0.2}

\appendixFigure{Model discovered using \texttt{IMf 0.3} on \textit{BPI Challenge 2019 (Sample of 3000 Cases)}}{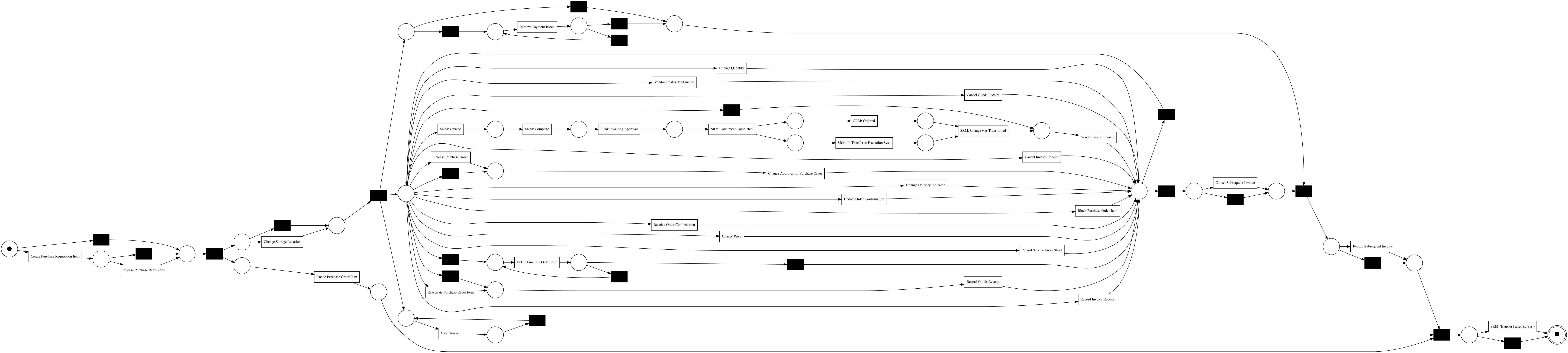}{appendix:fig:BPI_Challenge_2019_sampled_3000cases:IMf 0.3}

\appendixFigure{Model discovered using \texttt{IMf 0.4} on \textit{BPI Challenge 2019 (Sample of 3000 Cases)}}{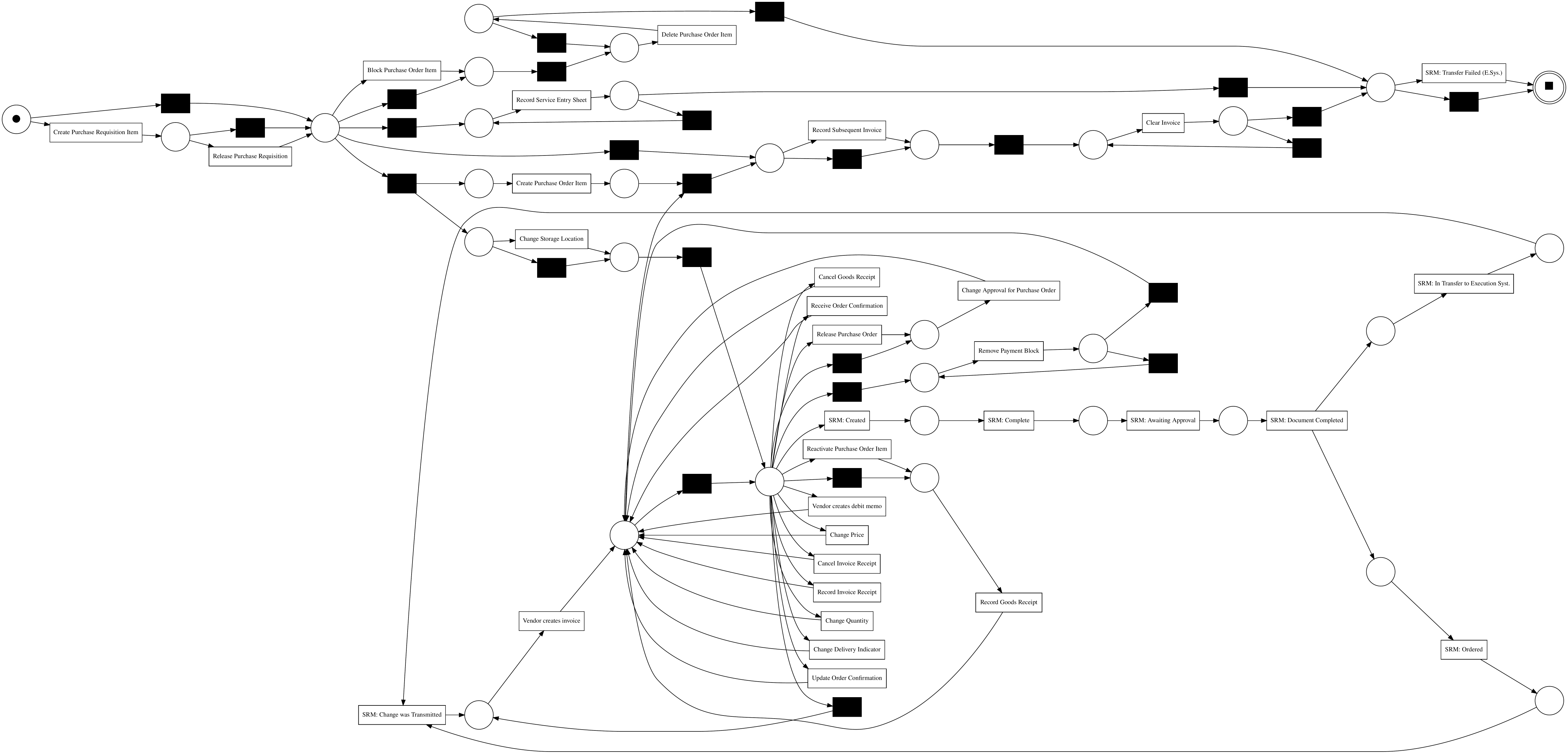}{appendix:fig:BPI_Challenge_2019_sampled_3000cases:IMf 0.4}

\appendixFigure{Model discovered using \texttt{$\alpha{}$ Top10} on \textit{BPI Challenge 2019 (Sample of 3000 Cases)}}{BPI_Challenge_2019_sampled_3000cases_model_alpha_Top10.pdf}{appendix:fig:BPI_Challenge_2019_sampled_3000cases:α Top10}

\appendixFigure{Model discovered using \texttt{$\alpha{}$ 10\%Cov} on \textit{BPI Challenge 2019 (Sample of 3000 Cases)}}{BPI_Challenge_2019_sampled_3000cases_model_alpha_10PERCENTCov.pdf}{appendix:fig:BPI_Challenge_2019_sampled_3000cases:α 10\%Cov}

\appendixFigure{Model discovered using \texttt{$\alpha{}$ 50\%Cov} on \textit{BPI Challenge 2019 (Sample of 3000 Cases)}}{BPI_Challenge_2019_sampled_3000cases_model_alpha_50PERCENTCov.pdf}{appendix:fig:BPI_Challenge_2019_sampled_3000cases:α 50\%Cov}

\appendixFigure{Model discovered using \texttt{$\alpha{}$ 80\%Cov} on \textit{BPI Challenge 2019 (Sample of 3000 Cases)}}{BPI_Challenge_2019_sampled_3000cases_model_alpha_80PERCENTCov.pdf}{appendix:fig:BPI_Challenge_2019_sampled_3000cases:α 80\%Cov}

\appendixFigure{Model discovered using \texttt{$\alpha{}$+++;2.0;b0.5;t0.5;r0.5} on \textit{BPI Challenge 2019 (Sample of 3000 Cases)}}{BPI_Challenge_2019_sampled_3000cases_model_alphappp-2.0-b0.5-t0.5-r0.5.pdf}{appendix:fig:BPI_Challenge_2019_sampled_3000cases:α+++|2.0|b0.5|t0.5|r0.5}

\appendixFigure{Model discovered using \texttt{$\alpha{}$+++;2.0;b0.3;t0.7;r0.6} on \textit{BPI Challenge 2019 (Sample of 3000 Cases)}}{BPI_Challenge_2019_sampled_3000cases_model_alphappp-2.0-b0.3-t0.7-r0.6.pdf}{appendix:fig:BPI_Challenge_2019_sampled_3000cases:α+++|2.0|b0.3|t0.7|r0.6}

\appendixFigure{Model discovered using \texttt{$\alpha{}$+++;2.0;b0.2;t0.8;r0.7} on \textit{BPI Challenge 2019 (Sample of 3000 Cases)}}{BPI_Challenge_2019_sampled_3000cases_model_alphappp-2.0-b0.2-t0.8-r0.7.pdf}{appendix:fig:BPI_Challenge_2019_sampled_3000cases:α+++|2.0|b0.2|t0.8|r0.7}

\appendixFigure{Model discovered using \texttt{$\alpha{}$+++;2.0;b0.2;t0.8;r0.8} on \textit{BPI Challenge 2019 (Sample of 3000 Cases)}}{BPI_Challenge_2019_sampled_3000cases_model_alphappp-2.0-b0.2-t0.8-r0.8.pdf}{appendix:fig:BPI_Challenge_2019_sampled_3000cases:α+++|2.0|b0.2|t0.8|r0.8}

\appendixFigure{Model discovered using \texttt{$\alpha{}$+++;2.0;b0.1;t0.9;r0.9} on \textit{BPI Challenge 2019 (Sample of 3000 Cases)}}{BPI_Challenge_2019_sampled_3000cases_model_alphappp-2.0-b0.1-t0.9-r0.9.pdf}{appendix:fig:BPI_Challenge_2019_sampled_3000cases:α+++|2.0|b0.1|t0.9|r0.9}

\appendixFigure{Model discovered using \texttt{$\alpha{}$+++;4.0;b0.5;t0.5;r0.5} on \textit{BPI Challenge 2019 (Sample of 3000 Cases)}}{BPI_Challenge_2019_sampled_3000cases_model_alphappp-4.0-b0.5-t0.5-r0.5.pdf}{appendix:fig:BPI_Challenge_2019_sampled_3000cases:α+++|4.0|b0.5|t0.5|r0.5}

\appendixFigure{Model discovered using \texttt{$\alpha{}$+++;4.0;b0.3;t0.7;r0.6} on \textit{BPI Challenge 2019 (Sample of 3000 Cases)}}{BPI_Challenge_2019_sampled_3000cases_model_alphappp-4.0-b0.3-t0.7-r0.6.pdf}{appendix:fig:BPI_Challenge_2019_sampled_3000cases:α+++|4.0|b0.3|t0.7|r0.6}

\appendixFigure{Model discovered using \texttt{$\alpha{}$+++;4.0;b0.2;t0.8;r0.7} on \textit{BPI Challenge 2019 (Sample of 3000 Cases)}}{BPI_Challenge_2019_sampled_3000cases_model_alphappp-4.0-b0.2-t0.8-r0.7.pdf}{appendix:fig:BPI_Challenge_2019_sampled_3000cases:α+++|4.0|b0.2|t0.8|r0.7}

\appendixFigure{Model discovered using \texttt{$\alpha{}$+++;4.0;b0.2;t0.8;r0.8} on \textit{BPI Challenge 2019 (Sample of 3000 Cases)}}{BPI_Challenge_2019_sampled_3000cases_model_alphappp-4.0-b0.2-t0.8-r0.8.pdf}{appendix:fig:BPI_Challenge_2019_sampled_3000cases:α+++|4.0|b0.2|t0.8|r0.8}

\appendixFigure{Model discovered using \texttt{$\alpha{}$+++;4.0;b0.1;t0.9;r0.9} on \textit{BPI Challenge 2019 (Sample of 3000 Cases)}}{BPI_Challenge_2019_sampled_3000cases_model_alphappp-4.0-b0.1-t0.9-r0.9.pdf}{appendix:fig:BPI_Challenge_2019_sampled_3000cases:α+++|4.0|b0.1|t0.9|r0.9}

\clearpage
\subsection{BPI Challenge 2020 (Request for Payment)}
\label{appendix:log:BPI_Challenge_2020}
\appendixFigure{Model discovered using \texttt{IMf 0.1} on \textit{BPI Challenge 2020 (Request for Payment)}}{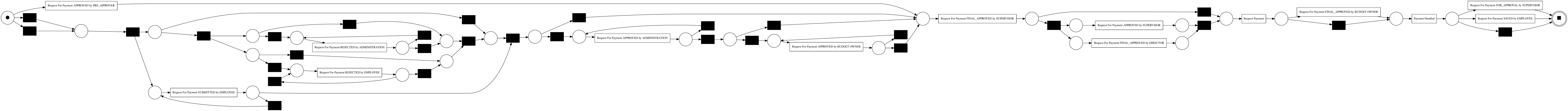}{appendix:fig:BPI_Challenge_2020:IMf 0.1}

\appendixFigure{Model discovered using \texttt{IMf 0.2} on \textit{BPI Challenge 2020 (Request for Payment)}}{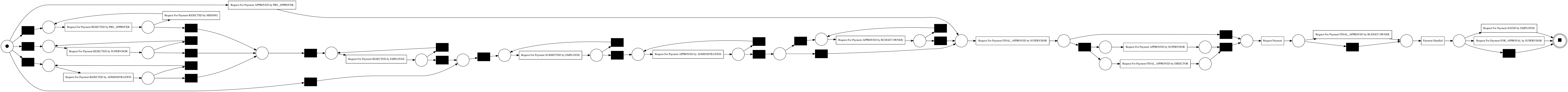}{appendix:fig:BPI_Challenge_2020:IMf 0.2}

\appendixFigure{Model discovered using \texttt{IMf 0.3} on \textit{BPI Challenge 2020 (Request for Payment)}}{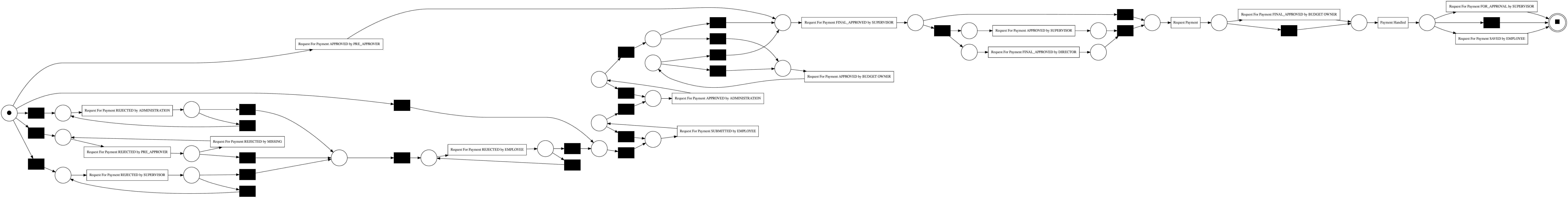}{appendix:fig:BPI_Challenge_2020:IMf 0.3}

\appendixFigure{Model discovered using \texttt{IMf 0.4} on \textit{BPI Challenge 2020 (Request for Payment)}}{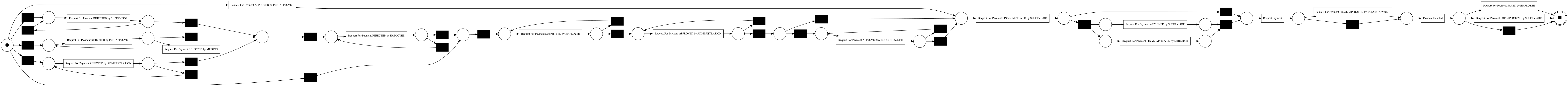}{appendix:fig:BPI_Challenge_2020:IMf 0.4}

\appendixFigure{Model discovered using \texttt{$\alpha{}$ Top10} on \textit{BPI Challenge 2020 (Request for Payment)}}{BPI_Challenge_2020_model_alpha_Top10.pdf}{appendix:fig:BPI_Challenge_2020:α Top10}

\appendixFigure{Model discovered using \texttt{$\alpha{}$ 10\%Cov} on \textit{BPI Challenge 2020 (Request for Payment)}}{BPI_Challenge_2020_model_alpha_10PERCENTCov.pdf}{appendix:fig:BPI_Challenge_2020:α 10\%Cov}

\appendixFigure{Model discovered using \texttt{$\alpha{}$ 50\%Cov} on \textit{BPI Challenge 2020 (Request for Payment)}}{BPI_Challenge_2020_model_alpha_50PERCENTCov.pdf}{appendix:fig:BPI_Challenge_2020:α 50\%Cov}

\appendixFigure{Model discovered using \texttt{$\alpha{}$ 80\%Cov} on \textit{BPI Challenge 2020 (Request for Payment)}}{BPI_Challenge_2020_model_alpha_80PERCENTCov.pdf}{appendix:fig:BPI_Challenge_2020:α 80\%Cov}

\appendixFigure{Model discovered using \texttt{$\alpha{}$+++;2.0;b0.5;t0.5;r0.5} on \textit{BPI Challenge 2020 (Request for Payment)}}{BPI_Challenge_2020_model_alphappp-2.0-b0.5-t0.5-r0.5.pdf}{appendix:fig:BPI_Challenge_2020:α+++|2.0|b0.5|t0.5|r0.5}

\appendixFigure{Model discovered using \texttt{$\alpha{}$+++;2.0;b0.3;t0.7;r0.6} on \textit{BPI Challenge 2020 (Request for Payment)}}{BPI_Challenge_2020_model_alphappp-2.0-b0.3-t0.7-r0.6.pdf}{appendix:fig:BPI_Challenge_2020:α+++|2.0|b0.3|t0.7|r0.6}

\appendixFigure{Model discovered using \texttt{$\alpha{}$+++;2.0;b0.2;t0.8;r0.7} on \textit{BPI Challenge 2020 (Request for Payment)}}{BPI_Challenge_2020_model_alphappp-2.0-b0.2-t0.8-r0.7.pdf}{appendix:fig:BPI_Challenge_2020:α+++|2.0|b0.2|t0.8|r0.7}

\appendixFigure{Model discovered using \texttt{$\alpha{}$+++;2.0;b0.2;t0.8;r0.8} on \textit{BPI Challenge 2020 (Request for Payment)}}{BPI_Challenge_2020_model_alphappp-2.0-b0.2-t0.8-r0.8.pdf}{appendix:fig:BPI_Challenge_2020:α+++|2.0|b0.2|t0.8|r0.8}

\appendixFigure{Model discovered using \texttt{$\alpha{}$+++;2.0;b0.1;t0.9;r0.9} on \textit{BPI Challenge 2020 (Request for Payment)}}{BPI_Challenge_2020_model_alphappp-2.0-b0.1-t0.9-r0.9.pdf}{appendix:fig:BPI_Challenge_2020:α+++|2.0|b0.1|t0.9|r0.9}

\appendixFigure{Model discovered using \texttt{$\alpha{}$+++;4.0;b0.5;t0.5;r0.5} on \textit{BPI Challenge 2020 (Request for Payment)}}{BPI_Challenge_2020_model_alphappp-4.0-b0.5-t0.5-r0.5.pdf}{appendix:fig:BPI_Challenge_2020:α+++|4.0|b0.5|t0.5|r0.5}

\appendixFigure{Model discovered using \texttt{$\alpha{}$+++;4.0;b0.3;t0.7;r0.6} on \textit{BPI Challenge 2020 (Request for Payment)}}{BPI_Challenge_2020_model_alphappp-4.0-b0.3-t0.7-r0.6.pdf}{appendix:fig:BPI_Challenge_2020:α+++|4.0|b0.3|t0.7|r0.6}

\appendixFigure{Model discovered using \texttt{$\alpha{}$+++;4.0;b0.2;t0.8;r0.7} on \textit{BPI Challenge 2020 (Request for Payment)}}{BPI_Challenge_2020_model_alphappp-4.0-b0.2-t0.8-r0.7.pdf}{appendix:fig:BPI_Challenge_2020:α+++|4.0|b0.2|t0.8|r0.7}

\appendixFigure{Model discovered using \texttt{$\alpha{}$+++;4.0;b0.2;t0.8;r0.8} on \textit{BPI Challenge 2020 (Request for Payment)}}{BPI_Challenge_2020_model_alphappp-4.0-b0.2-t0.8-r0.8.pdf}{appendix:fig:BPI_Challenge_2020:α+++|4.0|b0.2|t0.8|r0.8}

\appendixFigure{Model discovered using \texttt{$\alpha{}$+++;4.0;b0.1;t0.9;r0.9} on \textit{BPI Challenge 2020 (Request for Payment)}}{BPI_Challenge_2020_model_alphappp-4.0-b0.1-t0.9-r0.9.pdf}{appendix:fig:BPI_Challenge_2020:α+++|4.0|b0.1|t0.9|r0.9}

\clearpage
\subsection{BPI Challenge 2020 (Domestic Declarations)}
\label{appendix:log:BPI_Challenge_2020_DomesticDeclarations}
\appendixFigure{Model discovered using \texttt{IMf 0.1} on \textit{BPI Challenge 2020 (Domestic Declarations)}}{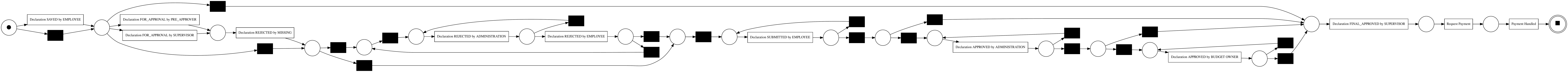}{appendix:fig:BPI_Challenge_2020_DomesticDeclarations:IMf 0.1}

\appendixFigure{Model discovered using \texttt{IMf 0.2} on \textit{BPI Challenge 2020 (Domestic Declarations)}}{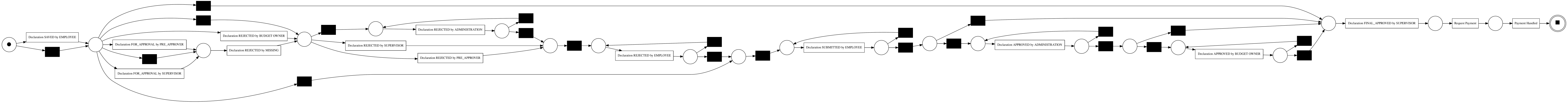}{appendix:fig:BPI_Challenge_2020_DomesticDeclarations:IMf 0.2}

\appendixFigure{Model discovered using \texttt{IMf 0.3} on \textit{BPI Challenge 2020 (Domestic Declarations)}}{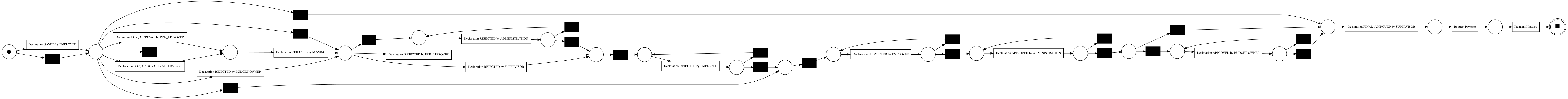}{appendix:fig:BPI_Challenge_2020_DomesticDeclarations:IMf 0.3}

\appendixFigure{Model discovered using \texttt{IMf 0.4} on \textit{BPI Challenge 2020 (Domestic Declarations)}}{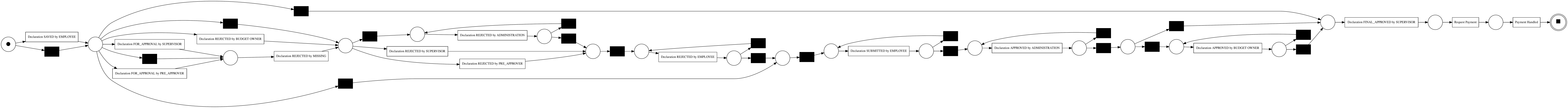}{appendix:fig:BPI_Challenge_2020_DomesticDeclarations:IMf 0.4}

\appendixFigure{Model discovered using \texttt{$\alpha{}$ Top10} on \textit{BPI Challenge 2020 (Domestic Declarations)}}{BPI_Challenge_2020_DomesticDeclarations_model_alpha_Top10.pdf}{appendix:fig:BPI_Challenge_2020_DomesticDeclarations:α Top10}

\appendixFigure{Model discovered using \texttt{$\alpha{}$ 10\%Cov} on \textit{BPI Challenge 2020 (Domestic Declarations)}}{BPI_Challenge_2020_DomesticDeclarations_model_alpha_10PERCENTCov.pdf}{appendix:fig:BPI_Challenge_2020_DomesticDeclarations:α 10\%Cov}

\appendixFigure{Model discovered using \texttt{$\alpha{}$ 50\%Cov} on \textit{BPI Challenge 2020 (Domestic Declarations)}}{BPI_Challenge_2020_DomesticDeclarations_model_alpha_50PERCENTCov.pdf}{appendix:fig:BPI_Challenge_2020_DomesticDeclarations:α 50\%Cov}

\appendixFigure{Model discovered using \texttt{$\alpha{}$ 80\%Cov} on \textit{BPI Challenge 2020 (Domestic Declarations)}}{BPI_Challenge_2020_DomesticDeclarations_model_alpha_80PERCENTCov.pdf}{appendix:fig:BPI_Challenge_2020_DomesticDeclarations:α 80\%Cov}

\appendixFigure{Model discovered using \texttt{$\alpha{}$+++;2.0;b0.5;t0.5;r0.5} on \textit{BPI Challenge 2020 (Domestic Declarations)}}{BPI_Challenge_2020_DomesticDeclarations_model_alphappp-2.0-b0.5-t0.5-r0.5.pdf}{appendix:fig:BPI_Challenge_2020_DomesticDeclarations:α+++|2.0|b0.5|t0.5|r0.5}

\appendixFigure{Model discovered using \texttt{$\alpha{}$+++;2.0;b0.3;t0.7;r0.6} on \textit{BPI Challenge 2020 (Domestic Declarations)}}{BPI_Challenge_2020_DomesticDeclarations_model_alphappp-2.0-b0.3-t0.7-r0.6.pdf}{appendix:fig:BPI_Challenge_2020_DomesticDeclarations:α+++|2.0|b0.3|t0.7|r0.6}

\appendixFigure{Model discovered using \texttt{$\alpha{}$+++;2.0;b0.2;t0.8;r0.7} on \textit{BPI Challenge 2020 (Domestic Declarations)}}{BPI_Challenge_2020_DomesticDeclarations_model_alphappp-2.0-b0.2-t0.8-r0.7.pdf}{appendix:fig:BPI_Challenge_2020_DomesticDeclarations:α+++|2.0|b0.2|t0.8|r0.7}

\appendixFigure{Model discovered using \texttt{$\alpha{}$+++;2.0;b0.2;t0.8;r0.8} on \textit{BPI Challenge 2020 (Domestic Declarations)}}{BPI_Challenge_2020_DomesticDeclarations_model_alphappp-2.0-b0.2-t0.8-r0.8.pdf}{appendix:fig:BPI_Challenge_2020_DomesticDeclarations:α+++|2.0|b0.2|t0.8|r0.8}

\appendixFigure{Model discovered using \texttt{$\alpha{}$+++;2.0;b0.1;t0.9;r0.9} on \textit{BPI Challenge 2020 (Domestic Declarations)}}{BPI_Challenge_2020_DomesticDeclarations_model_alphappp-2.0-b0.1-t0.9-r0.9.pdf}{appendix:fig:BPI_Challenge_2020_DomesticDeclarations:α+++|2.0|b0.1|t0.9|r0.9}

\appendixFigure{Model discovered using \texttt{$\alpha{}$+++;4.0;b0.5;t0.5;r0.5} on \textit{BPI Challenge 2020 (Domestic Declarations)}}{BPI_Challenge_2020_DomesticDeclarations_model_alphappp-4.0-b0.5-t0.5-r0.5.pdf}{appendix:fig:BPI_Challenge_2020_DomesticDeclarations:α+++|4.0|b0.5|t0.5|r0.5}

\appendixFigure{Model discovered using \texttt{$\alpha{}$+++;4.0;b0.3;t0.7;r0.6} on \textit{BPI Challenge 2020 (Domestic Declarations)}}{BPI_Challenge_2020_DomesticDeclarations_model_alphappp-4.0-b0.3-t0.7-r0.6.pdf}{appendix:fig:BPI_Challenge_2020_DomesticDeclarations:α+++|4.0|b0.3|t0.7|r0.6}

\appendixFigure{Model discovered using \texttt{$\alpha{}$+++;4.0;b0.2;t0.8;r0.7} on \textit{BPI Challenge 2020 (Domestic Declarations)}}{BPI_Challenge_2020_DomesticDeclarations_model_alphappp-4.0-b0.2-t0.8-r0.7.pdf}{appendix:fig:BPI_Challenge_2020_DomesticDeclarations:α+++|4.0|b0.2|t0.8|r0.7}

\appendixFigure{Model discovered using \texttt{$\alpha{}$+++;4.0;b0.2;t0.8;r0.8} on \textit{BPI Challenge 2020 (Domestic Declarations)}}{BPI_Challenge_2020_DomesticDeclarations_model_alphappp-4.0-b0.2-t0.8-r0.8.pdf}{appendix:fig:BPI_Challenge_2020_DomesticDeclarations:α+++|4.0|b0.2|t0.8|r0.8}

\appendixFigure{Model discovered using \texttt{$\alpha{}$+++;4.0;b0.1;t0.9;r0.9} on \textit{BPI Challenge 2020 (Domestic Declarations)}}{BPI_Challenge_2020_DomesticDeclarations_model_alphappp-4.0-b0.1-t0.9-r0.9.pdf}{appendix:fig:BPI_Challenge_2020_DomesticDeclarations:α+++|4.0|b0.1|t0.9|r0.9}

%% file: appendix-simplicity-generalization-table.tex
We additionally evaluated all the Petri nets discovered for the evaluation in \autoref{sec:eval} in terms of simplicity and generalization.
For that, we again utilized PM4Py using the corresponding evaluation functions.
In those, simplicity is measured using the inverse arc degree of nodes in the Petri net, while generalization relies on how frequent model elements are revisited during replay\footnote{See also \url{https://pm4py.fit.fraunhofer.de/documentation\#evaluation}.}.
\newcolumntype{X}{>{\centering\let\newline\\\arraybackslash\hspace{0pt}}b{1.2cm}}
\begin{sidewaystable}[htbp]
	\centering
	\caption{Simplicity and Generalization Results}
	\scalebox{0.8}{
		\begin{tabular}{l|XXXX|XXXX|XXXXXXXXXX|}
			\cline{2-19}
			                                                                                                                                                                             &
			\multicolumn{4}{c|}{\textbf{Inductive Miner Infrequent}}                                                                                                                     &
			\multicolumn{4}{c|}{\textbf{Alpha Miner}}                                                                                                                                    &
			\multicolumn{10}{c|}{\textbf{Alpha+++ Algorithm}}                                                                                                                              \\ \cline{2-19}
			                                                                                                                                                                             &
			\multicolumn{4}{c|}{Noise Threshold}                                                                                                                                         &
			\multicolumn{4}{c|}{Variant Filtering}                                                                                                                                       &
			\multicolumn{5}{c|}{\textit{Artificial Activity Threshold of 2.0}}                                                                                                           &
			\multicolumn{5}{c|}{\textit{Artificial Activity Threshold of 4.0}}                                                                                                             \\
			                                                                                                                                                                             &
			0.2                                                                                                                                                                          &
			0.3                                                                                                                                                                          &
			0.4                                                                                                                                                                          &
			0.5                                                                                                                                                                          &
			Top10                                                                                                                                                                        &
			10\%                                                                                                                                                                         &
			50\%                                                                                                                                                                         &
			80\%                                                                                                                                                                         &
			\begin{tabular}[c]{@{}c@{}}$b{=}0.5$\\ $t{=}0.5$\\ $r{=}0.5$\end{tabular}                         &
			\begin{tabular}[c]{@{}c@{}}$b{=}0.3$\\ $t{=}0.7$\\ $r{=}0.6$\end{tabular}                         &
			\begin{tabular}[c]{@{}c@{}}$b{=}0.2$\\ $t{=}0.8$\\ $r{=}0.7$\end{tabular}                      &
			\begin{tabular}[c]{@{}c@{}}$b{=}0.2$\\ $t{=}0.8$\\ $r{=}0.8$\end{tabular}                      &
			\multicolumn{1}{c|}{\begin{tabular}[c]{@{}c@{}}$b{=}0.1$\\ $t{=}0.9$\\ $r{=}0.9$\end{tabular}} &
			\begin{tabular}[c]{@{}c@{}}$b{=}0.5$\\ $t{=}0.5$\\ $r{=}0.5$\end{tabular}                      &
			\begin{tabular}[c]{@{}c@{}}$b{=}0.3$\\ $t{=}0.7$\\ $r{=}0.6$\end{tabular}                      &
			\begin{tabular}[c]{@{}c@{}}$b{=}0.2$\\ $t{=}0.8$\\ $r{=}0.7$\end{tabular}                      &
			\begin{tabular}[c]{@{}c@{}}$b{=}0.2$\\ $t{=}0.8$\\ $r{=}0.8$\end{tabular}                      &
			\begin{tabular}[c]{@{}c@{}}$b{=}0.1$\\ $t{=}0.9$\\ $r{=}0.9$\end{tabular}                        \\ \hline
			\multicolumn{1}{|l|}{\textbf{RTFM}}                                                                                                                                          &
			                                                                                                                                                                             &
			                                                                                                                                                                             &
			                                                                                                                                                                             &
			                                                                                                                                                                             &
			                                                                                                                                                                             &
			                                                                                                                                                                             &
			                                                                                                                                                                             &
			                                                                                                                                                                             &
			                                                                                                                                                                             &
			                                                                                                                                                                             &
			                                                                                                                                                                             &
			                                                                                                                                                                             &
			\multicolumn{1}{c|}{}                                                                                                                                                        &
			                                                                                                                                                                             &
			                                                                                                                                                                             &
			                                                                                                                                                                             &
			                                                                                                                                                                             &
			\\
			\multicolumn{1}{|l|}{Simplicity}                                                                                                                                             &
			0.6098                                                                                                                                                                       &
			0.6098                                                                                                                                                                       &
			0.6000                                                                                                                                                                       &
			0.7021                                                                                                                                                                       &
			1.0000                                                                                                                                                                       &
			1.0000                                                                                                                                                                       &
			1.0000                                                                                                                                                                       &
			1.0000                                                                                                                                                                       &
			0.7273                                                                                                                                                                       &
			0.8286                                                                                                                                                                       &
			1.0000                                                                                                                                                                       &
			1.0000                                                                                                                                                                       &
			\multicolumn{1}{c|}{1.0000}                                                                                                                                                  &
			0.7561                                                                                                                                                                       &
			0.8750                                                                                                                                                                       &
			1.0000                                                                                                                                                                       &
			1.0000                                                                                                                                                                       &
			1.0000                                                                                                                                                                         \\
			\multicolumn{1}{|l|}{Generalization}                                                                                                                                         &
			0.7299                                                                                                                                                                       &
			0.9844                                                                                                                                                                       &
			0.9792                                                                                                                                                                       &
			0.9343                                                                                                                                                                       &
			0.9936                                                                                                                                                                       &
			0.9966                                                                                                                                                                       &
			0.9966                                                                                                                                                                       &
			0.9966                                                                                                                                                                       &
			0.8645                                                                                                                                                                       &
			0.7398                                                                                                                                                                       &
			0.6776                                                                                                                                                                       &
			0.6776                                                                                                                                                                       &
			\multicolumn{1}{c|}{0.6776}                                                                                                                                                  &
			0.8645                                                                                                                                                                       &
			0.6776                                                                                                                                                                       &
			0.6776                                                                                                                                                                       &
			0.6776                                                                                                                                                                       &
			0.6776                                                                                                                                                                         \\ \hline
			\multicolumn{1}{|l|}{\textbf{Sepsis Cases}}                                                                                                                                  &
			                                                                                                                                                                             &
			                                                                                                                                                                             &
			                                                                                                                                                                             &
			                                                                                                                                                                             &
			                                                                                                                                                                             &
			                                                                                                                                                                             &
			                                                                                                                                                                             &
			                                                                                                                                                                             &
			                                                                                                                                                                             &
			                                                                                                                                                                             &
			                                                                                                                                                                             &
			                                                                                                                                                                             &
			\multicolumn{1}{c|}{}                                                                                                                                                        &
			                                                                                                                                                                             &
			                                                                                                                                                                             &
			                                                                                                                                                                             &
			                                                                                                                                                                             &
			\\
			\multicolumn{1}{|l|}{Simplicity}                                                                                                                                             &
			0.6183                                                                                                                                                                       &
			0.5929                                                                                                                                                                       &
			0.5914                                                                                                                                                                       &
			0.6190                                                                                                                                                                       &
			0.8462                                                                                                                                                                       &
			1.0000                                                                                                                                                                       &
			1.0000                                                                                                                                                                       &
			1.0000                                                                                                                                                                       &
			0.6364                                                                                                                                                                       &
			0.9500                                                                                                                                                                       &
			1.0000                                                                                                                                                                       &
			1.0000                                                                                                                                                                       &
			\multicolumn{1}{c|}{1.0000}                                                                                                                                                  &
			1.0000                                                                                                                                                                       &
			1.0000                                                                                                                                                                       &
			1.0000                                                                                                                                                                       &
			1.0000                                                                                                                                                                       &
			1.0000                                                                                                                                                                         \\
			\multicolumn{1}{|l|}{Generalization}                                                                                                                                         &
			0.8745                                                                                                                                                                       &
			0.8499                                                                                                                                                                       &
			0.8269                                                                                                                                                                       &
			0.7824                                                                                                                                                                       &
			0.9708                                                                                                                                                                       &
			0.9719                                                                                                                                                                       &
			0.9132                                                                                                                                                                       &
			0.9132                                                                                                                                                                       &
			0.8447                                                                                                                                                                       &
			0.8028                                                                                                                                                                       &
			0.7612                                                                                                                                                                       &
			0.6774                                                                                                                                                                       &
			\multicolumn{1}{c|}{0.6774}                                                                                                                                                  &
			0.9199                                                                                                                                                                       &
			0.8652                                                                                                                                                                       &
			0.8652                                                                                                                                                                       &
			0.8652                                                                                                                                                                       &
			0.8117                                                                                                                                                                         \\ \hline
			\multicolumn{1}{|l|}{\textbf{\begin{tabular}[c]{@{}l@{}}BPI Challenge 2019\\ (Sample of 3000 Cases)\end{tabular}}}                                                           &
			                                                                                                                                                                             &
			                                                                                                                                                                             &
			                                                                                                                                                                             &
			                                                                                                                                                                             &
			                                                                                                                                                                             &
			                                                                                                                                                                             &
			                                                                                                                                                                             &
			                                                                                                                                                                             &
			                                                                                                                                                                             &
			                                                                                                                                                                             &
			                                                                                                                                                                             &
			                                                                                                                                                                             &
			\multicolumn{1}{c|}{}                                                                                                                                                        &
			                                                                                                                                                                             &
			                                                                                                                                                                             &
			                                                                                                                                                                             &
			                                                                                                                                                                             &
			\\
			\multicolumn{1}{|l|}{Simplicity}                                                                                                                                             &
			0.6077                                                                                                                                                                       &
			0.5949                                                                                                                                                                       &
			0.5894                                                                                                                                                                       &
			0.6000                                                                                                                                                                       &
			0.6471                                                                                                                                                                       &
			1.0000                                                                                                                                                                       &
			0.8261                                                                                                                                                                       &
			0.4717                                                                                                                                                                       &
			0.7750                                                                                                                                                                       &
			1.0000                                                                                                                                                                       &
			1.0000                                                                                                                                                                       &
			1.0000                                                                                                                                                                       &
			\multicolumn{1}{c|}{1.0000}                                                                                                                                                  &
			0.6744                                                                                                                                                                       &
			0.8750                                                                                                                                                                       &
			1.0000                                                                                                                                                                       &
			1.0000                                                                                                                                                                       &
			1.0000                                                                                                                                                                         \\
			\multicolumn{1}{|l|}{Generalization}                                                                                                                                         &
			0.8183                                                                                                                                                                       &
			0.8442                                                                                                                                                                       &
			0.8372                                                                                                                                                                       &
			0.7833                                                                                                                                                                       &
			0.9634                                                                                                                                                                       &
			0.9811                                                                                                                                                                       &
			0.9716                                                                                                                                                                       &
			0.9590                                                                                                                                                                       &
			0.7469                                                                                                                                                                       &
			0.7469                                                                                                                                                                       &
			0.6975                                                                                                                                                                       &
			0.6975                                                                                                                                                                       &
			\multicolumn{1}{c|}{0.6900}                                                                                                                                                  &
			0.7617                                                                                                                                                                       &
			0.7617                                                                                                                                                                       &
			0.7617                                                                                                                                                                       &
			0.7617                                                                                                                                                                       &
			0.7284                                                                                                                                                                         \\ \hline
			\multicolumn{1}{|l|}{\textbf{\begin{tabular}[c]{@{}l@{}}BPI Challenge 2020\\ (Requests for Payment)\end{tabular}}}                                                           &
			                                                                                                                                                                             &
			                                                                                                                                                                             &
			                                                                                                                                                                             &
			                                                                                                                                                                             &
			                                                                                                                                                                             &
			                                                                                                                                                                             &
			                                                                                                                                                                             &
			                                                                                                                                                                             &
			                                                                                                                                                                             &
			                                                                                                                                                                             &
			                                                                                                                                                                             &
			                                                                                                                                                                             &
			\multicolumn{1}{c|}{}                                                                                                                                                        &
			                                                                                                                                                                             &
			                                                                                                                                                                             &
			                                                                                                                                                                             &
			                                                                                                                                                                             &
			\\
			\multicolumn{1}{|l|}{Simplicity}                                                                                                                                             &
			0.6727                                                                                                                                                                       &
			0.6923                                                                                                                                                                       &
			0.6923                                                                                                                                                                       &
			0.6923                                                                                                                                                                       &
			0.5610                                                                                                                                                                       &
			1.0000                                                                                                                                                                       &
			1.0000                                                                                                                                                                       &
			0.6923                                                                                                                                                                       &
			0.4773                                                                                                                                                                       &
			0.5882                                                                                                                                                                       &
			0.5882                                                                                                                                                                       &
			0.5882                                                                                                                                                                       &
			\multicolumn{1}{c|}{0.7255}                                                                                                                                                  &
			0.4651                                                                                                                                                                       &
			0.5758                                                                                                                                                                       &
			0.5758                                                                                                                                                                       &
			0.5758                                                                                                                                                                       &
			0.7143                                                                                                                                                                         \\
			\multicolumn{1}{|l|}{Generalization}                                                                                                                                         &
			0.8499                                                                                                                                                                       &
			0.7613                                                                                                                                                                       &
			0.7613                                                                                                                                                                       &
			0.7613                                                                                                                                                                       &
			0.9726                                                                                                                                                                       &
			0.9874                                                                                                                                                                       &
			0.9858                                                                                                                                                                       &
			0.9812                                                                                                                                                                       &
			0.8254                                                                                                                                                                       &
			0.8254                                                                                                                                                                       &
			0.8254                                                                                                                                                                       &
			0.8254                                                                                                                                                                       &
			\multicolumn{1}{c|}{0.7438}                                                                                                                                                  &
			0.8188                                                                                                                                                                       &
			0.8188                                                                                                                                                                       &
			0.8188                                                                                                                                                                       &
			0.8188                                                                                                                                                                       &
			0.7761                                                                                                                                                                         \\ \hline
			\multicolumn{1}{|l|}{\textbf{\begin{tabular}[c]{@{}l@{}}BPI Challenge 2020\\ (Domestic Declaration)\end{tabular}}}                                                           &
			                                                                                                                                                                             &
			                                                                                                                                                                             &
			                                                                                                                                                                             &
			                                                                                                                                                                             &
			                                                                                                                                                                             &
			                                                                                                                                                                             &
			                                                                                                                                                                             &
			                                                                                                                                                                             &
			                                                                                                                                                                             &
			                                                                                                                                                                             &
			                                                                                                                                                                             &
			                                                                                                                                                                             &
			\multicolumn{1}{c|}{}                                                                                                                                                        &
			                                                                                                                                                                             &
			                                                                                                                                                                             &
			                                                                                                                                                                             &
			                                                                                                                                                                             &
			\\
			\multicolumn{1}{|l|}{Simplicity}                                                                                                                                             &
			0.7067                                                                                                                                                                       &
			0.6667                                                                                                                                                                       &
			0.6667                                                                                                                                                                       &
			0.6667                                                                                                                                                                       &
			0.5217                                                                                                                                                                       &
			1.0000                                                                                                                                                                       &
			1.0000                                                                                                                                                                       &
			0.8750                                                                                                                                                                       &
			0.4699                                                                                                                                                                       &
			0.5278                                                                                                                                                                       &
			0.7200                                                                                                                                                                       &
			0.7200                                                                                                                                                                       &
			\multicolumn{1}{c|}{0.8919}                                                                                                                                                  &
			0.4500                                                                                                                                                                       &
			0.4737                                                                                                                                                                       &
			0.8421                                                                                                                                                                       &
			0.8421                                                                                                                                                                       &
			0.8857                                                                                                                                                                         \\
			\multicolumn{1}{|l|}{Generalization}                                                                                                                                         &
			0.8609                                                                                                                                                                       &
			0.7975                                                                                                                                                                       &
			0.7975                                                                                                                                                                       &
			0.7975                                                                                                                                                                       &
			0.9697                                                                                                                                                                       &
			0.9900                                                                                                                                                                       &
			0.9885                                                                                                                                                                       &
			0.9885                                                                                                                                                                       &
			0.8314                                                                                                                                                                       &
			0.8314                                                                                                                                                                       &
			0.8314                                                                                                                                                                       &
			0.8314                                                                                                                                                                       &
			\multicolumn{1}{c|}{0.7866}                                                                                                                                                  &
			0.8234                                                                                                                                                                       &
			0.8234                                                                                                                                                                       &
			0.8234                                                                                                                                                                       &
			0.8234                                                                                                                                                                       &
			0.8234                                                                                                                                                                         \\ \hline
		\end{tabular}
	}
\end{sidewaystable}